\documentclass[12pt]{iopart}
\usepackage{graphicx}
\usepackage{dcolumn}
\usepackage{bm}
\usepackage{braket}
\usepackage[usenames,dvipsnames]{xcolor}
\usepackage{caption}
\usepackage{subcaption}
\usepackage{bbold}
\usepackage{hyperref}
\usepackage{placeins}

\expandafter\let\csname equation*\endcsname\relax

\expandafter\let\csname endequation*\endcsname\relax

\usepackage{amsmath,amsthm, amssymb}
\graphicspath{{./Figure/}{./}}

\newcommand{\QQ}{\mathbb{Q}}

\newcommand{\ii}{ {\rm i} }

\newcommand{\ave}[1]{{\langle #1\rangle}}

\newcommand{\sL}{{\mathcal{L}}}

\usepackage{csquotes}
\newenvironment{itquote}
{\begin{quote}\itshape}
{\end{quote}}

\newcommand{\RR}{\mathbb{R}}

\begin{document}

\title[Non-stationarity and Dissipative Time Crystals]{Non-stationarity and Dissipative Time Crystals: Spectral Properties and Finite-Size Effects}

\author{Cameron Booker, Berislav Bu\v{c}a, Dieter Jaksch}

\address{Clarendon Laboratory, University of Oxford, Parks Road, Oxford OX1 3PU, United Kingdom}
\ead{berislav.buca@physics.ox.ac.uk}
\vspace{10pt}

\begin{abstract}
We discuss the emergence of non-stationarity in open quantum many-body systems. This leads us to the definition of dissipative time crystals which display experimentally observable, persistent, time-periodic oscillations induced by noisy contact with an environment. We use the Loschmidt echo and local observables to indicate the presence of a {finite sized} dissipative time crystal. Starting from the closed Hubbard model we then provide examples of dissipation mechanisms that yield experimentally observable quantum {periodic dynamics} and allow analysis of the emergence of {finite sized} dissipative time crystals. For a disordered Hubbard model including two-particle loss and gain we find a dark Hamiltonian driving oscillations between GHZ states in the long-time limit. Finally, we discuss how the presented examples could be experimentally realized.
\end{abstract}

%
%
\submitto{\NJP}
%
%
%

\section{Introduction}
Non-stationary many-body systems are those which cannot be described in terms of time-independent states or probability distributions. From seasonal weather patterns, across financial markets, to the behaviour of living organisms, such behaviour is ubiquitous in the world around us. Still, we lack a detailed understanding of how non-stationary, complex dynamics emerge from the microscopic laws governing the interactions between constituent quantum particles. This mystery was already noted in the seminal essay on complexity and symmetry ‘More is different’ \cite{Anderson} by P.W. Anderson who wrote:
\begin{itquote}
    Keeping on with the attempt to characterize types of broken symmetry which occur in living things, I find that at least one further phenomenon seems to be identifiable and either universal or remarkably common, namely, ordering (regularity or periodicity) in the time domain.
\end{itquote}

Statistical physics is arguably the most powerful physics approach to studying many-particle systems starting from microscopic quantum mechanical laws. However, even non-equilibrium statistical physics can only deal with probability distributions and quantum states that describe the system \emph{after} it has relaxed to stationarity. Relaxation to stationarity is widely assumed but the validity of this assumption has recently attracted significant theoretical and experimental scrutiny that we now briefly summarize.

\subsection{Relaxation to Stationarity in Many-body Quantum Systems}
For isolated systems the eigenstate thermalization hypothesis (ETH) \cite{ETHReview,ETHde,ETHalt} states that relaxation to stationarity happens by eigenstate dephasing \cite{eigenstatedephasing,EsslerGGE,thermoreview}. A physical quantity $\hat{O}$ at a time $t$ after evolving under the action of a Hamiltonian $H$ has expectation value
\begin{equation}
\ave{\hat{O}(t)}=\text{Tr} [\hat{O} \rho(t)]=\sum_{n,m}  e^{-\ii \omega_{nm} t} r_{nm} \bra{\phi_m} \hat{O}\ket{\phi_n}, \label{closeddyn}
\end{equation}
where $\ket{\phi_n}$ is an eigenstate of $H$ with energy $E_n$, and we have taken $\hbar = 1$. The coefficients $r_{nm}$ describe the initial state of the system $\rho(0)=\sum_{nm} r_{nm} \ket{\phi_n}\bra{\phi_m}$. 

The coherences in Eq. \eqref{closeddyn} oscillate with angular frequencies $\omega_{nm}=E_n-E_m$. In a generic many-body system the frequencies $\omega_{nm}$ are expected to be dense and incommensurate without any specific structure. The coherences will thus dephase quickly by destructive interference and take the system to stationarity within several characteristic times, e.g. a few hopping times for lattice models \cite{ETHReview}. In contrast, in real-world examples, such as those introduced above, non-stationarity persists for extremely long times compared to the characteristic atomic time-scales.

Exceptions to the behaviour predicted by the ETH in physical systems are mostly found in single-body low-temperature effects, such as solitions \cite{soliton}, or other low energy, single particle excitations \cite{Gabor,QPWT}. Another notable exception are quantum many-body scars \cite{scars,scars1,scars3,scarsexp,scars4,scarsdynsim,scardynsym2,scarsdynsym3,scarsdynsym4}. Here persistent non-stationarity, dubbed weak ergodicity breaking, has been linked to the existence of special eigenstates of $H$ that are almost equally spaced in energy and have high overlap with specific initial states with low-entanglement. Due to this, they exhibit persistent oscillations at one or few frequencies \cite{scars}. These scars are beginning to be understood through dynamical symmetries \cite{scarsdynsym4,scarsdynsim,scarsdynsym3,scardynsym2}.  Persistent oscillations have also been observed and studied in long-range interacting models \cite{Bojan1,Bojan2,longrange,Jad1,Jad2,Jad3}, models with confinement \cite{Confinement5,Confinement1,Confinement2,Confinement3,Confinement4,Confinement5,Confinement6} and \emph{many-body} breathing modes \cite{breathing1,breathing2,breathing3,MarkoJacopo,breathing4,breathing5,breathing6}. 

Similarly, in open quantum systems, one typically finds stationary behaviour at late times. Here, eigenstate dephasing is reduced since most modes of the system decay through interactions with the environment. Instead, system-environment interactions generally steer the many-body systems towards a unique, usually mixed, stationary state in the long-time limit. This is corroborated by random matrix theory calculations \cite{randommatrix1,randommatrix2,randommatrix3,randommatrix4,randommatrix5} and stationarity is implied when making the assumption of ergodicity.

Non-stationarity may arise in an open quantum system if it contains dark states \cite{BaumNarn,AlbertJiang, ChoiNoiselesssubsystems} which are eigenstates of the system Hamiltonian that are decoupled from the environment. An initial coherent superposition of dark states will continuously oscillate as if the system were isolated. Identifying dark states in many-body systems is challenging with only a limited number of examples, e.g. in a Heisenberg ring with ultra-local loss \cite{buca_non-stationary_2019} and driven dissipative Fermi-Hubbard models \cite{Yi_2012}, known.

\subsection{Origins of Time-Crystals}
\subsubsection{Closed and Periodically Driven Systems}
\hfill
\newline
\noindent Ordering in the time domain was studied by Wilczek in \cite{Wilczek} where he proposed quantum time crystals, which would be states of matter that spontaneously break time-translation symmetry, much like crystals break space translation symmetry and oscillate periodically. 
This original proposal was criticised on several grounds \cite{bruno1,bruno2}. Most notably, Watanabe and Oshikawa \cite{watanabe} formalised time crystalline behaviour as having an extensive observable, $\hat{O}$, which persistently oscillates at equilibrium with a single frequency $\omega$. They defined the equilibrium auto-correlation function,
\begin{equation}
 f(t) = \frac{\ave{\hat{O}(t)\hat{O}}_{eq}}{V^2}, \label{CF}
\end{equation}
where $V$ is the volume of the system. For a time crystal, at late times $f(t)$ would be time periodic, 
\begin{equation}
    f(t) \underset{t\rightarrow \infty}{=} C \cos(\omega t + \theta) + D, \label{WO}
\end{equation}
where $C,D$ are finite constants and $\theta$ is some phase. The motivation for this was to probe time-translation symmetry breaking in the same way that one probes space-translation symmetry breaking in crystals: as response functions at equilibrium. 
They then proved that, provided $H$ is sufficiently local, $C$ vanishes as $V\rightarrow\infty$, so $f(t)$ in fact becomes constant at late times \cite{watanabe}. However, even for equal time ($t=0$) correlators at finite temperature and a wide range of equilibrium states, $C$ vanishes in the thermodynamic limit \cite{Marko1}. In one-dimension, this is a consequence of the Mermin-Wanger theorem \cite{Cardybook}. The criterion proposed in \cite{watanabe} could, therefore, be interpreted as probing the absence of long-range spatial order rather than the presence of time order.

A  natural relaxation of the criterion \eqref{WO} is to consider a physically measurable correlation function,
\begin{equation}
\frac{\ave{\hat{O}(t)\hat{O}}_{eq}}{\ave{\hat{O}^2}_{eq}}\underset{t\to\infty}{=}C \cos{(\omega t + \theta)}+D, \label{MBJ}
\end{equation}
which is by construction normalized to $1$ at $t=0$. A finite, non-zero $C$ has been shown to be implied by the existence of a \emph{dynamical} extensive symmetry of the form \cite{buca_non-stationary_2019,Marko1,Marko2,chinzei2020time,scarsdynsim},
\begin{equation}
[H,A]=\omega A, \label{dynamicalsymmetry}
\end{equation}
where $A$ is an extensive operator with $\Tr(A^\dagger \hat{O})\neq 0$. 

Dynamical symmetries can result in the time-translation symmetry breaking of observables in closed systems following a quench, providing another possible definition of time crystalline behaviour. In the presence of a dynamical symmetry we assume the system to evolve towards a time-dependent generalized Gibbs ensemble (GGE)\footnote{This should be understood only as the effective state from which to calculate expectation values of local observables, as the actual state of the system remains pure.} \cite{Marko1,Marko2} given by
\begin{equation} \label{tdgge}
\rho(t)\underset{t\to\infty}{=} Z^{-1}\exp(- \beta H+\sum_j \mu^Q_j Q_j-\mu^A_j(t) A_j-\mu^A_j(t)^* A_j^\dagger),
\end{equation}
where $\beta$ is the temperature, $Z$ is a partition function, and $Q_j$ are conserved charges with $[H,Q_j] = 0$. The generalized chemical potentials $\mu_j^A(0)$ and $\mu_j^Q$ are determined from the initial state. We have allowed multiple orthogonal\footnote{In the Hilbert-Schmidt sense, $\langle A,B \rangle=\Tr{(A^\dagger B)}$} dynamical  symmetries $A_j$ with frequencies $\omega_j$ which determine the evolution of $\mu_j^A(t) = \mu_j^A(0) \exp(\ii \omega_j t)$. Equation \eqref{tdgge} is an extension of the well-established time-independent GGE that has been shown by both the ETH and extensive studies in integrable models to generically follow from a quantum quench in the absence of dynamical symmetries \cite{VidmarRigol}.

The presence of a large number of dense and incommensurate $\omega_j$ can lead to the absence of persistent oscillations and result in noisy dynamics or even dephasing to stationarity. This can be particularly important for free or many-body localized models \cite{thermoreview,VedikaReview} which are likely to have many $A_j$. We also remark that if the $A_j$ are not spatially translationally invariant this can lead to the absence of synchronization inside the system \cite{QuantumSynch,QuantSynch1,QuantSynch2}.

The quench-based time crystal definition has led to another particularly fruitful line of research, which is based on extending the notion of time translation symmetry breaking to discrete time translation symmetry breaking for Floquet systems with period $T$. More specifically, discrete time symmetry breaking is defined for local observables $o$ as a subharmonic response, $\bra{\psi}o(t+nT)\ket{\psi}=\bra{\psi}o(t)\ket{\psi}$, for some integer $n>1$, and $\bra{\psi}o(t+t_1)\ket{\psi}\neq\bra{\psi}o(t)\ket{\psi}$ for $ t_1<nT $ and for some generic enough initial state $\ket{\psi}$. This concept has been explored in both closed \cite{VedikaLazarides,timecrystal1,timecrystal2,timecrystal4,timecrystal5,timecrystal7,timecrystal8,timecrystal10,timecrystal11,timecrystal12} and open \cite{timecrystal6,timecrystal9, LazaridesRoy, timecrystal13} quantum systems. Various other extensions of the time crystal concept, including classical models \cite{classical1,classical3,classical4,Pablo,timecrystal7}, introducing non-local Hamiltonians \cite{kozin2019}, time quasi-crystals \cite{quasicrystal1} and time glass \cite{VedikaReview} have been proposed. 

For a more in depth review of the broad and exciting field of time crystals, the authors direct the reader to the excellent review article of Khemani et. al. \cite{VedikaReview}

\subsubsection{Open Systems}
\hfill
\newline
\noindent The works discussed above have been primarily focused on time crystals in closed and periodically driven systems. Realistically, physical systems are never perfectly isolated from their environment and in general one would expect system-environment interactions to disrupt the time crystalline behaviour of the previously discussed models. In this paper we will study extensions of quench based time crystals to open quantum systems which we label dissipative time crystals. These are characterized by persistent oscillations induced through the coupling with an external noisy environment \cite{buca_non-stationary_2019,Sarang,disTC,disTC2,disTC3,disTC4,Esslinger,Esslinger2,BucaJaksch2019,disTC5,disTC6,disTC7,disTC8,disTC9,disTC10,disTC11,disTC12,disTC13,disTC14,disTC15,disTC16}.  

In \cite{buca_non-stationary_2019} a general set of conditions leading to such behaviour was developed. These conditions were shown to be implied by the existence of strong dynamical symmetries. They were used \cite{BucaJaksch2019} to describe a recent experiment where a two-component Bose-Einstein condensate in a cavity, which is expected to exhibit chaotic behaviour and thermalization in the absence of dissipation, was stabilised up to a persistently oscillating phase by dissipation \cite{Esslinger,Esslinger2}.

Importantly, dissipative time crystals should be distinguished from related \emph{boundary} time crystals \cite{Fazio,boundary2} which are usually taken to be single-body systems that exhibit persistent oscillations despite dissipation. If these single-particle systems were closed they would exhibit simple Rabi oscillations (cf. \cite{chinzei2020time}) instead of thermalizing via the ETH.

In this paper, we will illustrate under what conditions coupling to an environment can induce time-crystalline behaviour. We will start from the Hubbard model as a genuine strongly correlated many-body system. In individual steps we will add two-body loss, two body gain and disorder to this system and explore their effect on the spectrum. Unlike in many-body localization-based time crystals, disorder will here not serve to stabilize the time crystal but rather introduce more different eigenfrequencies and increase the opportunity for eigenstate dephasing.

We will investigate the dynamics of these systems starting from a random pure initial state which we assume can be reliably created repeatedly to measure ensemble averages over different histories of interactions with the environment. This can e.g. be achieved in ultracold atom setups by using the same speckle pattern of pseudo-random on-site fields for each run of the experiment \cite{speckle}. The ensuing time evolution will be probed through the dynamics of local observables and the Loschmidt echo to detect dissipative time crystalline behaviour. The Loschmidt echo is widely used in many fields, e.g. quantum chaos \cite{Gorin} and more recently dynamical phase transitions \cite{DPTreview}. Its periodicity is one of the most stringent definitions for probing a time crystal because even for small systems the Poincare recurrence time diverges very fast with system size. As a result, so too does the period of the Loschmidt echo in the absence of criticality \cite{Zanardi,Zanardi2}. This should not be confused with revivals \cite{Echo1} for which a value of the Loschmidt echo close to the initial value 1 is restored aperiodically.

We will find that the noisy environment can indeed induce time crystalline behaviour despite the closed Hubbard model dephasing very quickly according to the ETH. We will show that the existence of a dissipative time crystal is closely related to dynamical symmetries but that they are neither necessary nor sufficient. For the case of a disordered Hubbard model with two-body loss, we furthermore derive a dark Hamiltonian \cite{buca_non-stationary_2019} describing the long-time coherent dynamics of the system. This non-local effective Hamiltonian is engineered from local physical dissipation and a local physical Hamiltonian rather than assumed outright (cf. \cite{kozin2019}).

The paper is organised as follows. In Secs. {\ref{sec:definitions}-\ref{sec:SDSs}} we define dissipative time crystals. More precisely, we outline the conditions required for them to arise {in finite systems} and demonstrate that these conditions are highly compatible with the existence of strong dynamical symmetries. In Sec. \ref{sec:hubbard} we study several non-stationarity probes for small 1D Hubbard models undergoing various types of dissipation in order to illustrate the conditions for which {finite sized} dissipative time crystals emerge. {In Sec. \ref{sec:infinite} we discuss the thermodynamic limit of our finite sized dissipative time crystals.} Finally, we conclude in Sec. \ref{sec:conc}.

\section{{Definitions of Dissipative Time Crystals}} \label{sec:definitions}
{We follow \cite{buca_non-stationary_2019} and define a dissipative time crystal as follows: a many-body quantum system coupled to a noise inducing environment which exhibits non-trivial periodic motion in some observable at late time for generic initial conditions. This naturally extends the notion of a time crystal as a closed quantum system which continuously oscillates periodically in time to the paradigm where the system can be disrupted by its environment.}

The key contrast with previous studies of time crystals is that here we are considering a system in the presence of noise, without any time-dependent driving. External noise/decoherence is generally understood as a mechanism for destroying quantum behaviour and inducing relaxation to stationarity, but here we explore the remarkable case where the external noise in fact induces persistent periodic motion within our system. This is of particular interest for uncovering experimentally realisable time crystals since in all practical settings the quantum systems we investigate are subject to some level of external noise from an environment that we cannot completely control. 

{In this work, we will primarily be interested in dissipative time crystals in finite systems. These are of particluar interest since they can be experimentally realised in cold atom experiments, as we will discuss later. We consider initialising a finite-sized, many-body system in a generic state and then require that after some finite transient time period during which parts of the system decay, the observable in question displays persistent, clearly measurable oscillations which do not decay further. We remark that these systems do not adhere to the view of \cite{Fazio} that a time crystal can appear only in the thermodynamic limit, however we will show that they can be easily modified so that the time translational symmetry breaking occurs only in this limit. Furthermore, as we will show, they demonstrate the remarkable phenomenon where external noise can stabilise the dynamics of an otherwise noisy system.}

{We will discuss the subtleties surrounding definitions of time crystals in open systems in more detail in Sec. \ref{sec:infinite}. Until then, unless stated otherwise, all discussions will be for strictly finite systems.}

\section{Spectral Requirements for Dissipative Time Crystals}\label{sec:dtc}

Under the assumptions that we have a sufficient handle on insulating our system from the environment so that the two are only weakly coupled and their interactions are Markovian, the natural framework in which to study these systems is the Lindblad master equation \cite{lindblad1976,BPTextbook,GardinerTextbook},
\begin{equation} \label{eq:masterequation}
    \begin{gathered}
        \dot{\rho}(t) = \mathcal{L}\rho = -\ii [H,\rho] + \sum_\mu \left( 2L_\mu \rho L_\mu^\dag -L_\mu^\dag L_\mu \rho - \rho L_\mu^\dag L_\mu \right).
    \end{gathered}
\end{equation}
This governs the evolution of a system with Hamiltonian $H$ whose coupling to the environment is described by the set of Lindblad jump operators $\{ L_\mu \}$. These operators model the influence of the environment's noise on our system. Such operators may for example be particle creation/annihilation operators to describe random particle gain/loss, or number operators to describe dephasing \cite{GardinerTextbook}. Through careful choice of which jump operators to include we can precisely characterise the effects of the system-environment interactions without a detailed description of the exact underlying mechanism.

The evolution of a system observable, $\hat{O}$, can then be described in terms of the eigensystem of the superoperator $\sL$. This is defined as the set $\{\rho_k, \sigma_k, \lambda_k\}$, where $\lambda_k$ are the eigenvalues of $\sL$ and $\rho_k, \sigma_k$ are the corresponding right and left eigenstates respectively. The eigensystem obeys the relations,
\begin{equation}
    \sL [\rho_k] = \lambda_k \rho_k, \ \sL^\dag [\sigma_k] = \lambda_k \sigma_k, \ \Tr(\sigma_k^\dag \rho_{k'}) = \delta_{k,k'}.
\end{equation}
Note that for physical $\sL$ the eigenvalues, $\lambda_k$, can only lie in the left half of the complex plane with $\text{Re}(\lambda_k) \le 0$.

The evolution of $\hat{O}$ given an initial state $\rho(0)$ can then be written as
\begin{equation} \label{eq:general_observable}
     \langle \hat{O} \rangle (t)  = \text{Tr}(\hat{O}\rho(t)) = \sum_k O_k e^{\lambda_k t},
\end{equation}
where $O_k = \Tr(\sigma_k^\dag \rho(0)) \Tr(\hat{O} \rho_k)$. We see immediately that in order to exhibit long time non-stationarity, there must exist some non-zero, purely imaginary eigenvalues, $\lambda_k = \ii \omega_k, \ \omega_k \ne 0$. In order for this non-stationarity to be periodic, the existence of these so-called quantum limit cycles \cite{limitcycle1,limitcycle2,disTC9} with purely imaginary eigenvalues is not sufficient, and instead we require that they be \emph{commensurable}, that is $\frac{\omega_k}{\omega_l} \in \QQ$ for all $k, l$. While this condition is theoretically sufficient for idealised dissipative time crystal behaviour, we explain later how in practice this must be modified slightly in order to yield a measurable dissipative time crystal.

Further, we can see that the transient time which we must wait before such time crystalline behaviour is visible is determined by the Liouvillian gap, $R$, defined as
\begin{equation}
    R = \min \{ \| \text{Re}(\lambda_k) \| : \text{Re}(\lambda_k)<0 \}.
\end{equation}
This characterises the time scales of the slowest decaying parts of the system.

As a result, it is clear that the study of dissipative time crystals in this regime is intrinsically linked to the study of the eigenvalues, especially the purely imaginary eigenvalues, of the Liouvillian superoperator $\sL$. We remark that the existence of commensurable, purely imaginary eigenvalues is highly non-trivial for Liouvillians describing many-body systems due to their exponential size. Consequently developing conditions under which these eigenvalues can exist in realisable Liouvillians is central to understanding dissipative time crystals.

\section{Dark States and Strong Dynamical Symmetries} \label{sec:SDSs}
As we have remarked above, in order to study dissipative time crystal we must understand the spectral properties of $\sL$. However, for many-body quantum systems, $\sL$ in general becomes exponentially complex and out of reach for current computational methods. To analytically find even all purely imaginary eigenvalues remains, in general, an open problem. However in the examples we have studied, we found them to arise only from two key mechanisms: dark states and strong dynamical symmetries.

Dark states \cite{Lidar,Albert}, defined by, 
\begin{equation}
    H\ket{\phi_n} = E_n \ket{\phi_n} , \ L_\mu \ket{\phi_n} = 0 \ \forall \mu,
\end{equation}
are eigenstates of the closed system Hamiltonian which are invisible to the external noise. They span a decoherence free subspace \cite{Lidar} which is left unaffected by the dissipation. 
The resulting dark state coherences, of the form $\rho_{n,m} = \ket{\phi_n}\bra{\phi_m}$, are then trivially eigenstates of $\sL$ with purely imaginary eigenvalues $\lambda_{n,m} = -\ii (E_n - E_m)$. We thus find that coherences between dark states with differing energies,  $E_n \ne E_m$, have the non-zero purely imaginary eigenvalues which are necessary for a dissipative time crystal. In a way, they may be understood as quantum scars \cite{scars,scars1,scarsdynsim,scarsdynsym3,scarsdynsym4,scardynsym2}, but now selected through the dissipation, instead of being special eigenstates.
 
In general, for large systems, the computation of these dark states and their energies remains a computationally unfeasible task, unless the system Hamiltonian has some integrability structure from which to find the energies, $E_n$, and eigenstates, $ \ket{\phi_n}$. Additionally, these must be suitably compatible with the jump operators $L_\mu$ such that the calculation of $L_\mu \ket{\phi_n}$ possible. We therefore look to more powerful symmetry arguments in order to deal with larger systems.

A strong dynamical symmetry, $A$, of the system with symmetry-frequency $\omega$ is defined by
\begin{equation}
    [H,A] = \omega A,\ [L_\mu, A] = [L_\mu^\dag, A] = 0, \ \omega \in \RR.
\end{equation}
While this trivially implies that $A$ is a  strong symmetry of $H$ with $[H, A] = \omega A$ \cite{BucaProsen,AlbertJiang,Zhao,BaumNarn}, note that the converse does not hold.
A theory of these symmetries has been studied and developed in the past \cite{buca_non-stationary_2019}. It has been shown that if a system has a strong dynamical symmetry $A$ and $\rho_\infty$ is a non-equilibrium steady state with $\sL [\rho_\infty] = 0$ (NESS), then mixed coherences of the form,
\begin{equation}
    \rho_{n,m} = (A^\dag)^n \rho_\infty A^m,
\end{equation}
are eigenstates of $\sL$ with purely imaginary eigenvalues, 
\begin{equation}
    \lambda_{n,m} = \ii(n-m)\omega .
\end{equation}
We also remark that it is possible for the system to have multiple independent symmetries $A_j$ with different symmetry-frequencies $\omega_j$. In this case we find several `strings' of imaginary eigenvalues, $\lambda_{n,m}^{j} = \ii (n-m) \omega_j$, which are integer multiples of the different $\omega_j$.

It is important to distinguish these mixed coherences from the decoherence free subspace of dark states discussed above. The dark states are completely decoupled from the dissipation through $L_\mu \ket{\phi_n} = 0$. Contrastingly, for mixed coherences we generally have $L_\mu \rho_{n,m} L_\mu^\dag \ne0$ which results in the time evolution being influenced by the dissipation.

While these two mechanisms allow us to show the existence of purely imaginary eigenvalues, unfortunately neither mechanism is in fact necessary or sufficient for time crystalline behaviour. In the case of having both a strong dynamical symmetry and purely imaginary eigenvalues for other reasons (e.g. dark states) the desired eigenvalue commensurability is generically broken. Alternatively, having multiple independent \emph{strong} dynamical symmetries can also break the desired spectral structure. We also demonstrate how commensurability can arise in the absence of any strong dynamical symmetries.

In the next section, we introduce examples that progressively reduce the number of eigenfrequencies in the asymptotic dynamics leading to increasingly more periodic behaviour. The least periodic one will be the closed model with an exponentially large number of eigenfrequencies, and we will reach a case with only a single pair of eigenfrequencies oscillating between two GHZ states. The focus will be on small systems to examine the finite-size effects which demonstrate the key points of the above discussion but we expect some results to be generalisable to larger systems. All examples are for physical local Hamiltonians and dissipation which we believe could in principle be experimentally realisable. 

\section{Dissipative Time Crystals in the 1D Hubbard Model} \label{sec:hubbard}
We now demonstrate the emergence of dissipative time crystals from the 1D Hubbard model by coupling it to an environment with two body loss and gain processes. These models are naturally realizable in cold atom settings \cite{coldexp}. We find that in fact, due to the exponentially large number of eigenfrequencies, we can illustrate the basic principles while looking at only very small systems.

\subsection{Methods}
The setup considered will be as follows: we initialise the system in some random pure state and then allow it to evolve while calculating the expectation values of two non-stationarity probes. Such a setup could in principle be realized in ultra-cold atoms by fixing once a choice of some very strong pseudo-random on-site fields, in addition to those already present in the model being studied, and then cooling the system so that it is (at least very nearly) in the ground state. This pure ground state then becomes our random pure initial state when we remove the additional fields and quench the system. Further, when considering a system coupled weakly to the environment, by using the same choice of pseudo-random on-site fields we can repeatedly construct the same initial state in order to construct ensemble averages.

The two non-stationarity probes which we shall consider are the transverse fermion spin on each site and the Loschmidt echo of the initial state. As per the definition in Sec. {\ref{sec:definitions}}, for the system to be a dissipative time crystal with respect to one of these probes, the probe's signal must persistently oscillate at later times.

The Loschmidt echo, defined as
\begin{equation}
    \mathcal{E}_\text{I} = \Tr(\rho(0)^\dag \rho(t)),
\end{equation}
is widely used in many fields, e.g. quantum chaos \cite{Gorin} and more recently dynamical phase transitions (e.g. \cite{DPTreview,Jad1,Jad2,Jad3}). For an initially pure state, $\rho(0) = \ket{\psi}\bra{\psi}$, it measures the fidelity between the current state of the system and the initial state. The periodicity of this is one of the most stringent definitions and probes of time crystal-like behaviour in our setup because it's evolution will typically be influenced by all the eigenfrequencies of $\sL$. Since there are generally a large number of eigenfrequencies in even small many-body systems, the Poincare recurrence time, and thus the period of the Loschmidt echo, can be effectively infinite for experimental purposes. Note, however, that the Loschmidt echo is not a practical probe in larger systems as the amplitude of the late time signal decays exponentially with system size. Nevertheless, in the examples we present below it excellently demonstrates the consequences of our previous discussion about the eigenvalue structure required for dissipative time crystals.

We use the discrete Fourier transform (DFT), denoted $F[f(t)]$, as a natural tool to detect oscillatory behaviour in the observables. All DFT spectra have been calculated using Blackman windowing to reduce spectral leakage caused by only measuring over a finite time period which does not necessarily contain an integer number of oscillations \cite{fftwindowing}. Further, we have computed all DFT's at late times after the transient contributions have vanished.

\subsection{Closed Hubbard Model} \label{sec:closed}
In order to show that the dissipative time crystal behaviour {in later examples} is not a trivial consequence of the dynamical properties and strong symmetry of the closed system, we first discuss the closed Hubbard model for spin 1/2 fermions on a chain of length $L$. The system is described by the Hamiltonian
\begin{equation}\label{eq:Hamiltonain}
    \begin{gathered}
        H = - \sum_{i=1}^{L-1} \sum_{s \in \{\uparrow, \downarrow\}} (c_{i,s}^\dag c_{i+1,s} + \text{h.c}) + \sum_{j=1}^{L} U_jn_{j,\uparrow}n_{j,\downarrow} \\ +\epsilon_j n_j + \frac{B}{2}(n_{j,\uparrow}-n_{j,\downarrow})
    \end{gathered}
\end{equation}
where $c_{j,s}$ annihilates a fermion of spin $s$ on site $j$ and obeys the canonical fermionic anti-commutation relations. The number operators are given by 
\begin{equation}
    n_{j,s} = c^\dag_{j,s}c_{j,s}, \ n_j = n_{j, \uparrow} + n_{j, \downarrow}.
\end{equation}
The Hamiltonian includes inhomogeneous onsite interactions $U_j$, an inhomogeneous spin independent potential $\epsilon_j$ and a uniform magnetic field $B$ aligned with the $z$-direction of the fermion spins. In all our numerics we chose the $U_j, \epsilon_j \in [0,3]$ to be a fixed realisation of i.i.d uniform random variables. The inhomogeneous interaction and potential break the usual $SU(2)$ $\eta$-symmetry of the system \cite{essler_frahm_gohmann_klumper_korepin_2005} so that we are left with only the spin $SU(2)$ symmetry,
\begin{equation}
    [H,S^z] = 0, \ [H,S^\pm] = \pm BS^\pm,    
\end{equation}
where 
\begin{equation}
    S^z = \sum_j \left[n_{j, \uparrow}-n_{j,\downarrow}\right], \ S^+ = \sum_j c_{j, \uparrow}^\dag c_{j, \downarrow}, \ S^+ = \sum_j c_{j, \downarrow}^\dag c_{j, \uparrow}.
\end{equation}
The transverse fermion spin is then given by,
\begin{equation}
    S_j^x = \tfrac12 (c^\dag_{j, \uparrow}c_{j, \downarrow} + c^\dag_{j, \downarrow}c_{j, \uparrow}).
\end{equation}

\begin{figure}[!htb]
    \centering
    \includegraphics[width = 0.75 \textwidth]{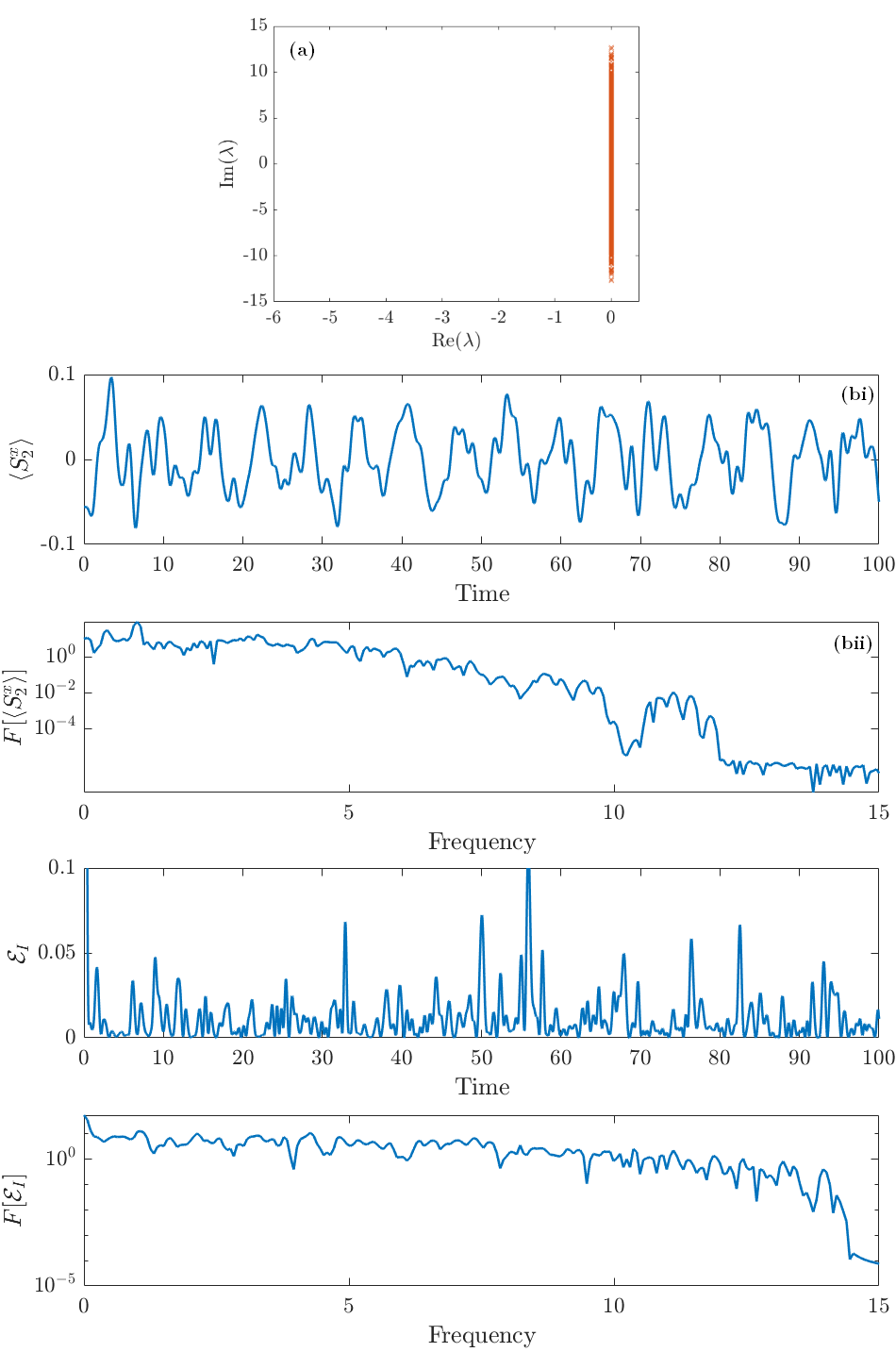}
    \caption{Closed System. (a) The Spectrum of a 3 site system, characteristic of larger systems. (b) \& (c) The evolution and DFT of the transverse spin on the second site and Loschmidt echo respectively for a 4 site system which is initialised in a random pure state.}
    \label{fig:closed_system}
\end{figure}

While the spectrum of $H$ (and thus $\sL$) can be found analytically for constant $U_j, \epsilon_j$ \cite{essler_frahm_gohmann_klumper_korepin_2005}, here we must resort to numerical calculations. We plot the spectrum of $\sL$ in Figure \ref{fig:closed_system}a for a $3$ site system. This system size is chosen so that the spectra may be directly compared across the different examples and in the open cases any larger systems are computationally out of reach. However it is reasonable to expect that the spectra of larger systems are qualitatively the same.  Importantly for the discussion of periodicity we find that the purely imaginary eigenvalues are \emph{effectively} incommensurable. By this we mean that while the numerically calculated eigenvalues are commensurable, as a result of finite precision computation, the time period over which any oscillations would occur is far longer than any reasonable time scale of an experiment. The immediate consequence of this is that unless we carefully choose our initial state and observable such that the expansion in Eq. \eqref{eq:general_observable} contains only a commensurable subset of frequencies, the signal will never be periodic on experimental timescales. In general, without such specific choices, the many incommensurate frequencies will interfere producing either noise in smaller systems or decay to stationarity by eigenstate dephasing \cite{eigenstatedephasing} in larger ones.

We can see the manifestation of this by looking at plots of the transverse spin and Loschmidt echo in Figures \ref{fig:closed_system}b and \ref{fig:closed_system}c where a 4 site system has been evolved from one instance of a random initial state $\ket{\phi} \propto \sum _n u_n\ket{\phi_n}$. The $u_n$ are i.i.d uniform random numbers on $[0,1]$ and $\ket{\phi_n}$ are the eigenstates of $H$. The calculations were repeated for many different random initial states to check that indeed the resulting behaviour is generic and not a consequence of some particular initial conditions. For both probes the signal is very noisy and far from periodic. The noise is suppressed slightly in the transverse spin signal since some of the eigenstates have no overlap with $S_k^x$, thus removing the corresponding frequencies in \eqref{eq:general_observable}. In addition, these observations are corroborated by the DFT spectra for the signal from both probes, which are roughly flat for the majority of the frequencies, characteristic of noise \cite{howard_signal_2015}.

We can thus conclude that in the absence of interactions with an environment, the 1D Hubbard model does not display periodic behaviour for generic initial conditions on experimental timescales even for very small systems. This can be understood by eigenstate dephasing and demonstrates that the {clearly periodic} behaviour we find in the open models is not a trivial consequence of the closed system's dynamics.

\subsection{Two-body Loss} \label{sec:loss}
The first dissipative case we consider is when the system leaks eta-pairs, each composed of a spin-up and a spin-down fermion, to the environment from every site at a possibly inhomogeneous but strictly positive rate, $\gamma_j>0$. This is described by introducing the set of jump operators $L_j = \gamma_j \eta_{j}^{-} = \gamma_j c_{j, \downarrow} c_{j, \uparrow}$ in Eq. \eqref{eq:masterequation} while keeping the same Hamiltonian as in Eq. \eqref{eq:Hamiltonain}. 

In this system we find numerically that the only purely imaginary eigenvalues correspond to coherences between dark states which are left unaffected by the loss. These dark states have no overlap with any eta pairs.  We expect that one could prove the non-existence of other purely imaginary eigenvalues as a consequence of Sup. Thm. 2 of \cite{buca_non-stationary_2019_sup}, but this is beyond the scope of the current work. There exists a strong dynamical symmetry, generated by $S^+$ with $[H, S^\pm] = \pm B S^\pm$. However, we do not find any mixed coherences here as all NESSs are pure dark states and so the states of the form $(S^+)^n \rho_\infty (S^-)^m$ are coherences between dark states. All states with purely imaginary eigenvalues are independent of the $\gamma_j$, and so the only relevance of these coupling strengths for our discussion is to determine the time period of the transient dynamics. Thus in what follows we will take $\gamma_j = 1$ for all sites.

\begin{figure}[!htb]
    \centering
    \includegraphics[width = 0.75 \textwidth]{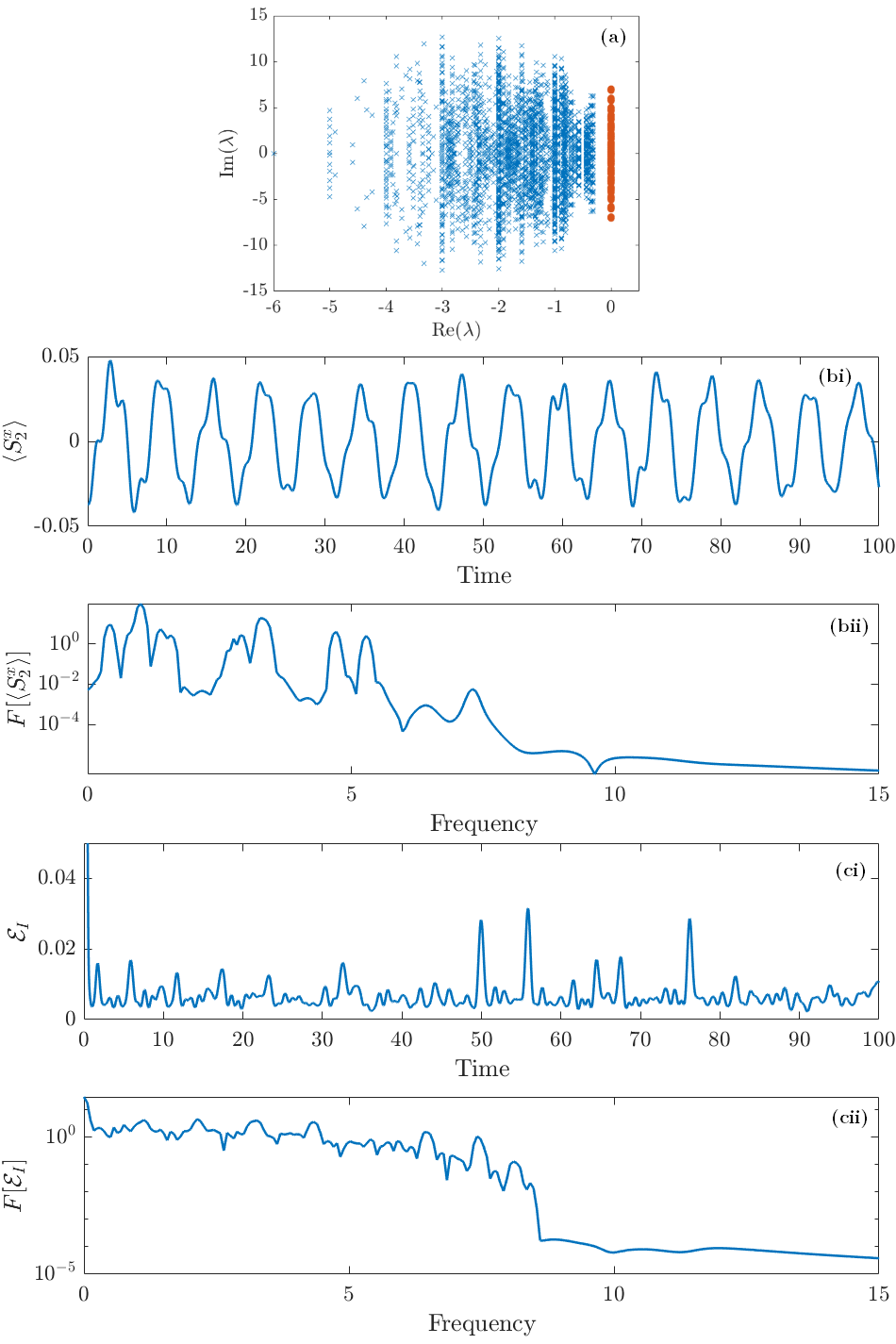}
    \caption{Open system with pure loss. (a) The Spectrum of a 3 site system, characteristic of larger systems which are computationally out of reach. The purely imaginary eigenvalue have been highlighted in red. (b) \& (c) The evolution and DFT of the transverse spin on the second site and Loschmidt echo respectively for a 4 site system which is initialised in a random pure state.}
    \label{fig:loss_system}
\end{figure}

In Figure \ref{fig:loss_system}a we plot the spectrum for a 3 site system, but again this is expected to be representative of larger systems. Comparing with the spectrum of the closed system, we see that there are significantly fewer imaginary eigenvalues. Again, however, we find that the remaining imaginary eigenvalues are effectively incommensurate. 

We examine the time dynamics of our two stationarity probes in Figures \ref{fig:loss_system}b \& c for a 4 site system in a random initial state. Starting with the echo we see that the signal is extremely noisy, an observation which is corroborated by the DFT being roughly flat for much of the spectrum.

Looking next at the transverse spin signal we see that, while clearly not periodic, the signal is noticeably less noisy than for the closed case. As before this is a result of many coherences between dark states not having any overlap with $S_k^x$ and thus the series in Eq. \eqref{eq:general_observable} having significantly fewer terms. This is clear from the DFT spectrum where we can now see the emergence of roughly 8-10 pronounced peaks at effectively incommensurable frequencies. While this is not strictly time crystalline behaviour, it can reasonably be called quasi-time crystalline behaviour, in analogy to quasi-periodicity created by a small number of incommensurable frequencies \cite{quasicrystal1,quasicrystal2}.

Through exploring this example, we have demonstrated that in the presence of only two body losses, our 1D Hubbard model does not exhibit experimentally observable dissipative time crystalline behaviour owing again to the effectively incommensurable eigenfrequencies. This shows how the existence of a strong dynamical symmetry is not sufficient for such behaviour as this does not guarantee commensurable purely imaginary eigenvalues. We have however found evidence of quasi-time crystalline behaviour, at least with respect to the transverse spin, $S^x_k$. This weaker condition requires only that there are few frequencies present in the evolution, not that they are commensurable. We do not explore quasi-time crystal behaviour any further here, however we envisage this to be an interesting area for further study.

\subsection{Two-body loss and gain} \label{sec:lossgain}
To finally uncover truly dissipative time crystalline behaviour, we now additionally introduce two body gain terms into the Liouvillian \eqref{eq:masterequation}, as $L_j = \Gamma_j \eta_j^+ = \Gamma_j c_{j, \uparrow}^\dag c_{j, \downarrow}^\dag$, again for $\Gamma_j>0$. This now represents a system where eta-pairs are both gained and lost on each site at various rates. Importantly these additional jump operators do not break the strong dynamical $S^+$ symmetry.

With both two body loss and gain, we can prove that the space of dark states is restricted such that they are exactly the states
\begin{equation}
    \ket{\phi_n} = (S^+)^n \ket{\downarrow \downarrow\dots \downarrow}.
\end{equation}
Coherences between these states, $\ket{\phi_n} \bra{\phi_m}$, are eigenstates of $\sL$ with purely imaginary eigenvalues $\lambda_{n,m} = -\ii(n-m)B$. We see that these are in fact generated by the strong dynamical symmetry since they can all be written as
\begin{equation}
    \ket{\phi_n} \bra{\phi_m} = (S^+)^n \rho_\text{down}(S^-)^m,
\end{equation}
where $\rho_\text{down} = \ket{\downarrow \downarrow\dots \downarrow}\bra{\downarrow \downarrow\dots \downarrow}$ is a NESS of $\sL$. We additionally find that this system also has mixed NESSs which are not trivial linear combinations of the $\ket{\phi_n} \bra{\phi_n}$. Consequently the strong dynamical symmetry also generates true mixed coherences which are distinct from the space of dark states but also have eigenvalues $\lambda_{n,m} = -\ii(n-m)B$. Numerically, it was found that there are no further states with purely imaginary eigenvalues, and so for this system \emph{all} non-decaying eigenstates are generated by the strong dynamical $S^+$ symmetry.

\begin{figure}[!htb]
    \centering
    \includegraphics[width = 0.75 \textwidth]{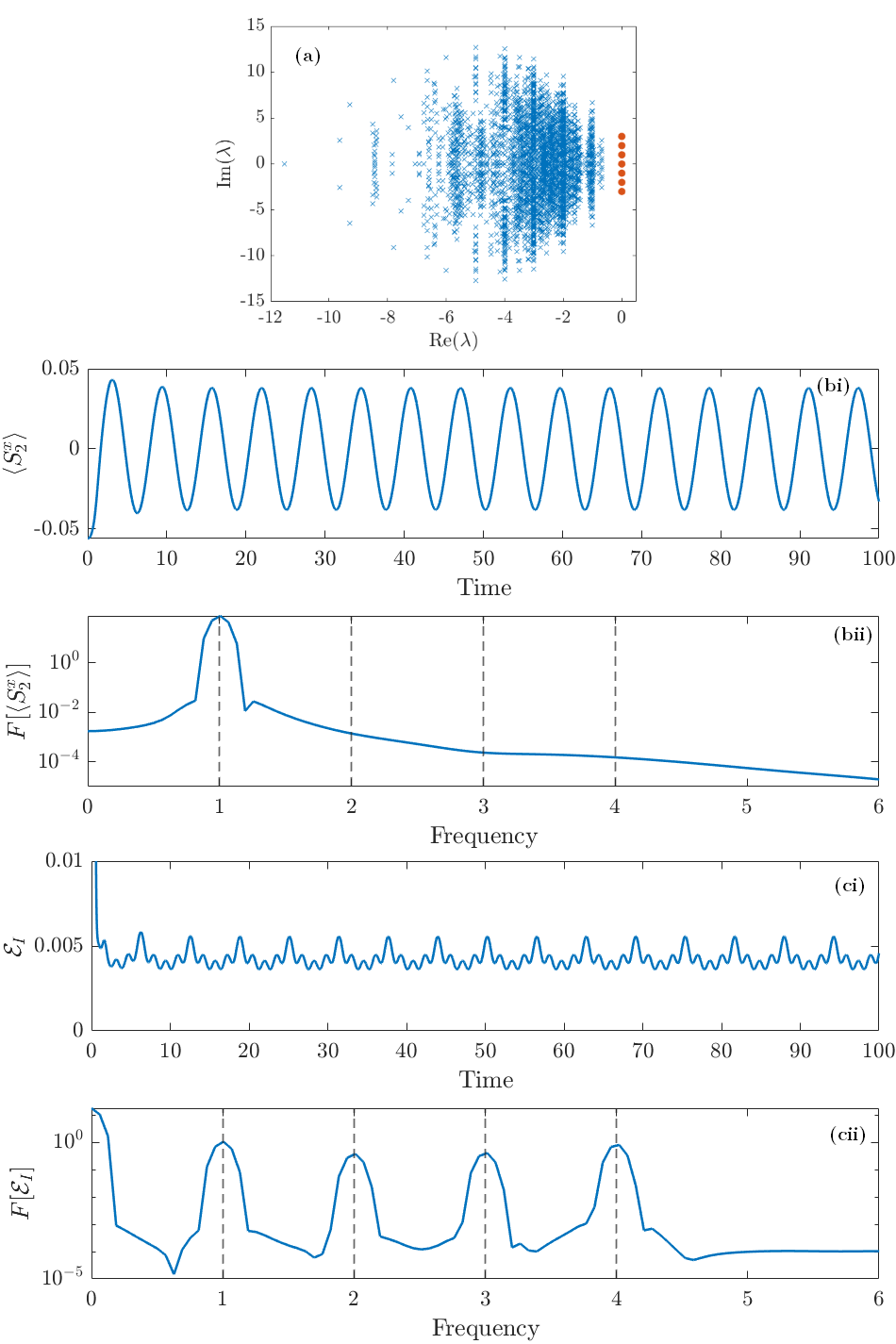}
    \caption{Open system with both loss and gain. (a) The Spectrum of a 3 site system, characteristic of larger systems which are computationally out of reach. The purely imaginary eigenvalues have been highlighted in red. (b) \& (c) The evolution and DFT of the transverse spin on the second site and Loschmidt echo respectively for a 4 site system which is initialised in a random pure state. We highlight the non-zero imaginary eigenvalues by dashed vertical lines on the spectra.}
    \label{fig:lossgain_system}
\end{figure}

As a consequence of the strong dynamical $S^+$ symmetry generating all the purely imaginary eigenvalues, they are all integer multiples of $B$ and thus easily seen to be commensurable. We show this in Figure \ref{fig:lossgain_system}a for  $\gamma_j = \Gamma_j = 1$, $\forall j$. This is an example of the eigenvalue structure that we require for time crystalline behaviour. 

Indeed when we now look at the evolution of our probes (Figures \ref{fig:lossgain_system}b \& c) we see that both the transverse spin and Loschmidt echo rapidly decay into persistent oscillations. For the echo, the DTF spectrum shows that all purely imaginary eigenvalues are present whereas for the transverse spin only the eigenstates with imaginary eigenvalue $\pm \ii B$ contribute.

Here we have found the emergence of a dissipative time crystal. Critically the purely imaginary eigenvalues have the necessary structure of commensurability that is required for periodic behaviour. We have further seen that this structure has been provided by the strong dynamical symmetry generating \emph{all} non-decaying eigenstates.

\begin{figure}[h!]
    \centering
    \includegraphics[width = 0.75 \columnwidth]{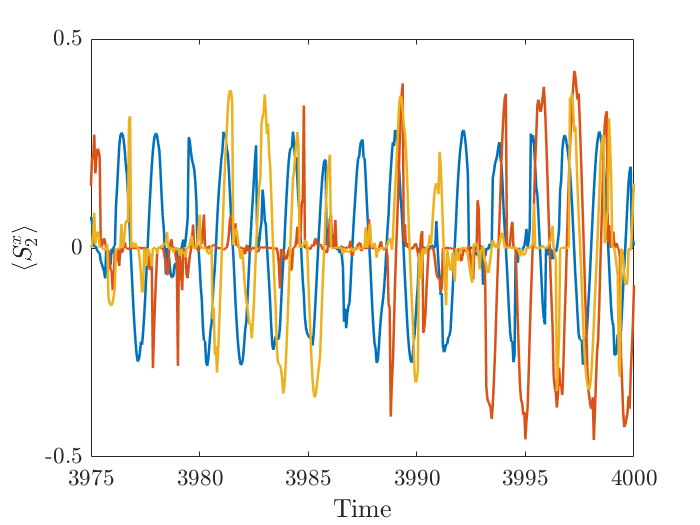}
    \caption{Three quantum trajectories of the same initial pure state for the system in Sec. \ref{sec:lossgain} with both two body loss and gain. In a 4 site system, we calculate the transverse spin on site 2 at very late times after transient dynamics have vanished to demonstrate that each trajectory is different.}
    \label{fig:lossgaintrajectories}
\end{figure}

We also note that, counterintuitively, as we can see in Figure \ref{fig:lossgaintrajectories}, each individual quantum trajectory \cite{DaleyQuantumTrajectories} for the same initial state (obtained through stochastic unravelling of the master equations) is different, even at very late times after transient behaviour has vanished. This essentially embodies the stochastic nature of the quantum bath that acts on the system (cf. with \cite{LazaridesRoy}). This may also be understood as a non-stationary quantum stochastic process \cite{Nurdin_2020}.

\subsection{Inhomogeneous Magnetic Field} \label{sec:mag}
\begin{figure}[!htb]
    \centering
    \includegraphics[width = 0.75 \textwidth]{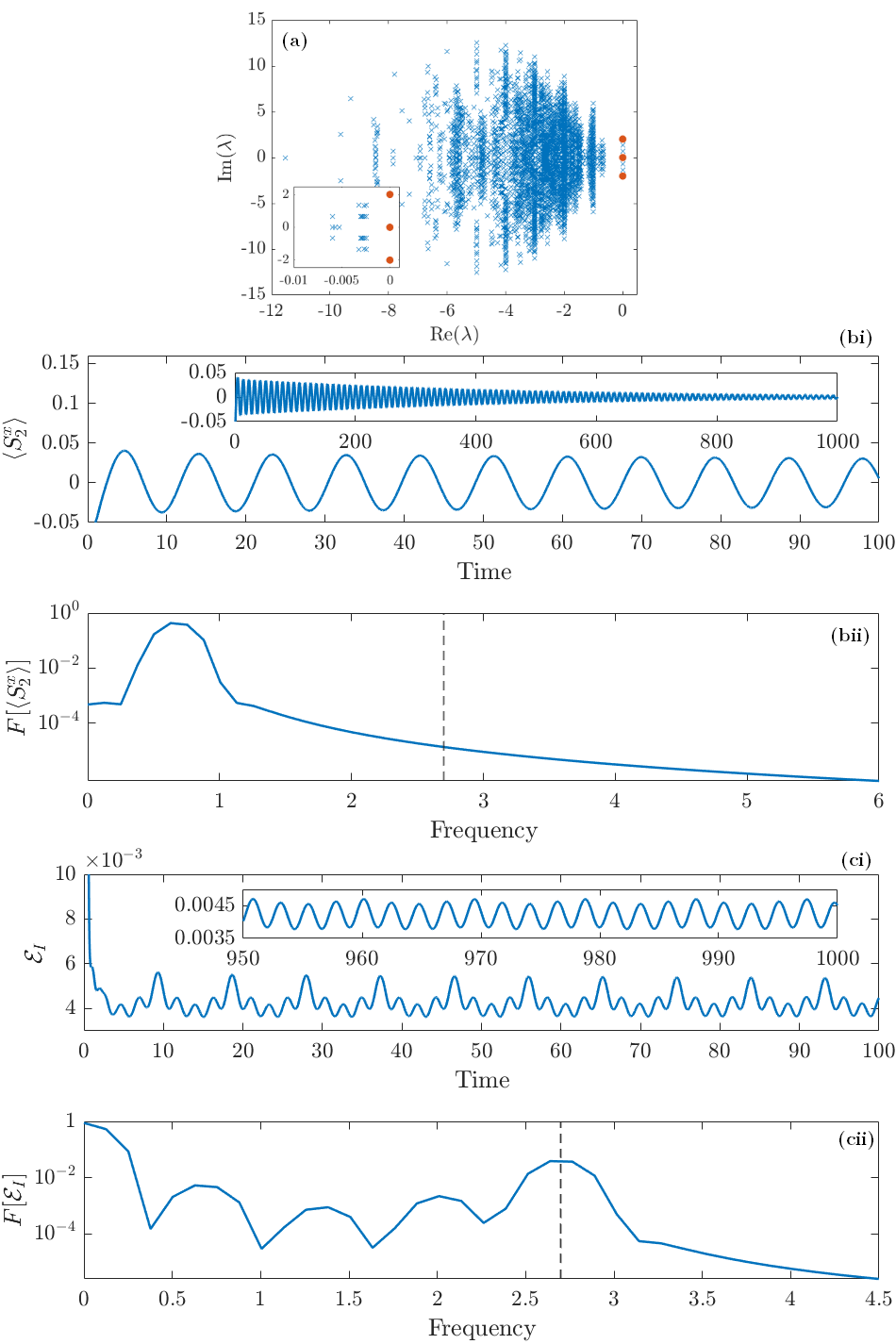}
    \caption{Open system with disordered magnetic field with both loss and gain. (a) The Spectrum of a 3 site system, characteristic of larger systems which are computationally out of reach. The purely imaginary eigenvalues have been highlighted in red. The inset focuses on the eigenvalues with very small but non-zero real part. (b) \& (c) The evolution and DFT of the transverse spin on the second site and Loschmidt echo respectively for a 4 site system which is initialised in a random pure state. We highlight the non-zero purely imaginary eigenvalue $\mathcal{B}$ with a dashed vertical line on the spectra. In this case we have calculated both DFTs for the range $950< t< 1000$ as shown in the inset of (ci). }
    \label{fig:ghz_system}
\end{figure}

Finally, we take the system above with both two-body loss and gain but introduce a weakly inhomogeneous magnetic field, $B_j$, in the Hamiltonian \eqref{eq:Hamiltonain}. This breaks the strong dynamical $S^+$ symmetry in addition to the structure of the dark states. As a result, the two remaining dark states are the all-spin-up and all-spin-down states and further we find that coherences between these two states are the only eigenstates of $\sL$ with non-zero, purely imaginary eigenvalues, given by $\pm \ii \sum_j B_j$. At late times the system therefore can only oscillate at a single frequency. In Figure \ref{fig:ghz_system}a we also see that there are several eigenvalues with very small real parts, corresponding to states which would have purely imaginary eigenvalues if the disorder in $B_j$ were removed, resulting in a very small Liouvillian gap. 

The consequence of the very small Liovillian gap is immediately seen in the signals of our two probes. It is easy to see that the coherences between all-spin-up and all-spin-down states have no overlap with $S_k^x$ for any $k$ and so after the transient period we have $S^x_k = \text{const}$. Thus comparing Figure \ref{fig:ghz_system}b with the previous systems we find that here it takes at least an order of magnitude longer for the transient behaviour to die away. We observe the same behaviour in the echo (Figure \ref{fig:ghz_system}c), and the inset demonstrates that at sufficiently late times, once the transient modes have become almost insignificant, the system reaches persistent oscillations at the single frequency we expect.

After the transient period, the dynamics of the system on top of the steady states can be described by the effective, non-local Hamiltonian,
\begin{equation}\label{eq:ghz_Heff}
    H_\text{eff} = \frac{\mathcal{B}}{2} \Big( \ket{\uparrow\dots\uparrow}\bra{\uparrow\dots\uparrow} - \ket{\downarrow\dots \downarrow}\bra{\downarrow\dots\downarrow} \Big)
\end{equation}
where $\mathcal{B} = \sum_j B_j$. This drives coherent oscillations between two GHZ states,
\begin{equation}
    \ket{\text{GHZ}}_\pm = \frac{\ket{\uparrow\dots \uparrow} \pm \ket{\downarrow \dots \downarrow}}{\sqrt{2}}.
\end{equation}
Note that here we have engineered an effective non-local dark Hamiltonian \cite{buca_non-stationary_2019} through a local and physical Hamiltonian in the presence of noisy dissipation, rather than assuming a closed non-local Hamiltonian with unclear stability properties \cite{kozin2019,Vedikacomment,Eisert}. However, the amplitude of the resulting dynamics depends on the overlap between the initial states and the GHZ states, which for arbitrary initial states will decrease exponentially with system size. 

We see from this model dissipative time crystal{line} behaviour in the absence of any strong dynamical symmetries and thus conclude that such symmetries are indeed neither necessary nor sufficient. 

\FloatBarrier

\subsection{Discussion}

These examples demonstrate how dissipative time crystalline behaviour can be induced in a finite-sized system, starting from the 1D Hubbard model through noisy contact with an external environment. They also show how purely imaginary eigenvalues of $\sL$ are not alone sufficient for such behaviour to emerge. Instead we see that these eigenvalues must at least be commensurable. 

In fact, commensurability alone is not sufficient spectral structure in general. For a robust dissipative time crystal in a practical setting we desire persistent oscillations that are \emph{clearly} measurable. We discussed in Sec \ref{sec:closed} that as a result of finite precision all frequencies in both numerical calculations and experimental measurements will be commensurable. However, we may find the corresponding time periods are significantly longer than any experimental time scales and so the motion is effectively aperiodic.
We thus introduce the notion of \emph{effective} incommensurability as commensurable frequencies with a resultant time period longer than the experimental time scale. Another way in which commensurable frequencies can become effectively incommensurable is if they are too dense with respect to the inverse of the experimental timescale since then the time period can grow exponentially in the number of frequencies. In such a case, the spectral peaks of the signals DFT could also become merged and indistinguishable, rendering the commensurability of the frequencies and hence the presence of persistent periodic motion possibly undetectable. While the examples studied numerically here do not exhibit such behaviour, since quantum many-body systems have exponentially many eigenvalues it is in principal feasible for this to occur and thus should be considered in future works.

This condition of nowhere-dense, commensurable, purely imaginary eigenvalues which are not effectively incommensurable is in general highly non-trivial for Liouvillians of many-body systems. Nevertheless it can be guaranteed by the existence of a single strong dynamical symmetry \cite{buca_non-stationary_2019} which generates all continuously oscillating modes. Further, we can easily see that the imaginary eigenvalues will not be closely spaced provided the symmetry-frequency is not too small with respect to the inverse of the experimental timescale. In experiment, this should be avoidable as the symmetry frequency can generally be tuned by experimental parameters, but it remains an important consideration nonetheless. If instead there are purely imaginary eigenvalues beyond those guaranteed by the strong dynamical symmetry, e.g. coming from dark states, we see that having a strong dynamical symmetry is not sufficient for time crystalline behaviour. This is illustrated by the example having two-body loss in Sec. \ref{sec:loss}. We can further envisage the existence of two distinct strong dynamical symmetries with effectively incommensurate frequencies and easily see that again such a system would not be periodic for generic initial conditions and observables on experimental timescales. In contrast, the example in Sec. \ref{sec:mag} with an inhomogeneous magnetic field demonstrates that strong dynamical symmetries are not necessary for dissipative time crystals.

While these examples have been focused on finite-sized systems, the qualitative properties can be expected to hold more generally in certain larger systems. More specifically, the persistent oscillations in an observable $\hat{O}$ are a result of the expansion
\begin{equation}
    \langle \hat{O} \rangle (t)  = \sum_k O_k e^{-\ii \omega_k t}, \ O_k = \Tr(\sigma_k^\dag \rho(0)) \Tr(\hat{O} \rho_k),
\end{equation}
containing only commensurable frequencies $\omega_k$. For the Loschmidt echo in our our examples, we find that the coefficients $O_k$ decay exponentially with system size. However, by conservation of the total spin, $S^2 = (S^x)^2 + (S^y)^2 + (S^z)^2$ we see that the same cannot occur for $S_k^x$ and so we expect that these persistent oscillations would be present in larger systems.

\section{{From Finite to Inifinite Dissipative Time Crystals}} \label{sec:infinite}
{When traditionally discussing time crystals in closed systems with unitary dynamics, the time translational symmetry breaking must occur in the thermodynamic limit in order to exclude the trivial oscillations of finite quantum systems. As a consequence of this, it is sometimes felt that even in the open case the spontanious time translational symmetry breaking should occur \emph{only} in the thermodynamic limit \cite{Fazio}. This means that in a finite system the symmetry breaking will only be transient and meta-stable. We believe this requirement to be too restictive since it excludes many interesting situations where non-trivial time translational symmetry breaking can occur in open systems. In particular it exludes the highly non-trivial finite size dissipative time crystals discussed above, where the noisy dynamics of a many-body system can be remarkably stabilised by noisy interactions with an external bath. It also exludes many experimental platforms, such as ultracold atoms, where only finite systems can be studied. It is for these reasons that we use the term dissipative time crystal in its most general sense of \emph{any} dissipation induced non-trivial time translational symmetry breaking.
}

{However, one can modify our example in Sec. \ref{sec:lossgain} with both loss and gain so that the time translational symmetry breaking occurs strictly in the limit of infinite system size. To do so we simply add an additional local Lindblad operator which breaks the strong dynamical symmetry, say $L = \mu n_1$, to the Liouvillian. For any finite system size there are now no non-zero purely imaginary eigenvalues due to lack of a strong dynamical symmetry. However, as a consequence of finite velocity of propagation \cite{Lieb_Robinson_1972} the effect of the ultra-local term becomes irrelevant in the bulk of the thermodynamically large system (see also \cite{Znidaric_2015}). Therefore, only the thermodynamically large system has purely imaginary eigenvalues and exhibits time-translational symmetry breaking. More generally, this argument applies to any finite size dissipative time crystal where the purely imaginary eigenvalues are the consequence of a strong dynamical symmetry.}

{This provides additional interest for the dissipative time crystals which we have discussed in detail. Not only are they remarkable in their own right as finite system which exhibit intriguing, experimentally realisable physics, they can also be easily extended to the more restrictive notion of a time crystal as a thermodynamic phenomenon. We propose that this motivates further work to classify and define the plethora of interesting and contrasting non-trivial time translational symmetry breaking phenomena which can occur in these types of open quantum systems and can be encompassed by the general term dissipative time crystal.} 

\section{Conclusion and Outlook} \label{sec:conc}
We have reviewed the existing works on time crystals and extended the concept in order to define the dissipative time crystal in open quantum systems. Such systems are of interest as they exhibit persistent oscillations due to being in contact with a noisy environment. This is of particular interest for realisable time crystals where the system will never be completely isolated from noisy interactions with its surroundings. 

In the Markovian master equation framework, we have demonstrated the importance of structure within the purely imaginary eigenvalues of $\sL$ when looking to observe the breaking of time symmetry in open, many-body quantum systems. In particular the purely imaginary eigenvalues must form a nowhere-dense commensurable set which for robustness must not be effectively incommensurable. This is a highly non-trivial condition for general many-body Liouvillians which have exponentially many eigenvalues.

We have also introduced a strict measure of time-crystalline behaviour based on the periodicity of the Loschmidt echo of a generic initial state. This is a very stringent measure because even for very small many-body systems the time of periodicity (Poincare recurrence time) is expo-exponentially large in system size due to the exponential number of eigenfrequencies \cite{Zanardi} in particular in the absence of criticality \cite{Zanardi2}. We also consider \emph{random} initial states which will generally be expected to excite an exponentially large number of eigenfrequencies for a many-body system. Thus the time dissipative time crystal behaviour will be generally observed in the system and not a consequence of precisely engineered initial conditions and observables. We remark however that this can only probe finite-size effects on time-crystalization, as for a random initial state the amplitude of the Loschmidt echo is generally expected to decrease with system size

We looked at finite size examples based on the Hubbard model with two-body loss and gain, naturally realizable in cold atom platforms \cite{coldexp}. These allowed us to demonstrate the requirements of eigenvalue structure for dissipative time crystals and to explore how this non-trivial structure is related to mechanisms of dark states and strong dynamical symmetries. While the Loschmidt echo is not practically applicable to significantly larger systems, the dynamics of certain other observables are expected to remain qualitatively the same.

By considering first the isolated system we showed that the existence of purely imaginary eigenvalues of $\sL$ is not sufficient for persistent oscillations for generic initial conditions and observables. However, further examples illustrated that such behaviour can emerge provided the purely imaginary eigenvalues of the Liouvillian are at least commensurable. The example with both two-body loss and gain demonstrated that the condition of commensurability is guaranteed by the existence of a single strong dynamical symmetry provided it generates \emph{all} the continuously oscillating modes. In general, however, the existence of a strong dynamical symmetry is not sufficient for time crystalline behaviour, as illustrated by the example with only two-body loss. This example also allowed us to define a dissipative quasi-time crystal. Finally, the example with an inhomogeneous magnetic field provided an example of a dissipative time crystal without any dynamical symmetries. This model also intriguingly realizes an effective non-local \emph{dark Hamiltonian} at late times giving coherent dynamics between two GHZ states. Remarkably, such dynamics are here achieved with a purely local physical Hamiltonian and realistic dissipation and further are robust to variations in many of the model parameters. 

{We concluded by arguing that the dissipative time crystal which we examined in detail can be easily modified so as to adher to the sentiment that time crystals should be purely thermodynamic phenomena.}

Open questions still remain regarding a more general theory of time crystalline behaviour in {finite} open systems described by Lindblad master equations. In particular in exploring whether realistic Liouvillians of this type, describing many-body systems, can have additional non-zero, purely imaginary eigenvalues which are not described by dark states or strong symmetries. Additionally the example in Sec. \ref{sec:loss} suggests the study of dissipative \emph{quasi}-time crystals, where one can have only a few incommensurate frequencies in analogy with discrete time quasi-crystals \cite{quasicrystal1,quasicrystal2}. {Also, as discussed in Sec. \ref{sec:infinite}, much work is required to examine the subtleties that arise in the thermodynmic limit, and to carefully classify the range of dissipative time crystalline behaviour which can arrise.}

In future work we expect to study linear response theory of dissipative time crystals, behaviour under periodic driving, the formulation of a semiclassical limit \cite{Andreas}, and the influence of dissipation on persistent oscillations at quantum critical points \cite{Castro}, in gauge theories \cite{Confinement1,Delfino}, or low energy Luttinger liquid descriptions \cite{Luttinger1,Luttinger2}. Additionally the relations between our dynamical symmetry requirement under dissipation and those for quantum many-body scars will be explored \cite{scarsdynsim,scarsdynsym3,scarsdynsym4,scardynsym2}. {We also intend examine the thermodynamic limit of our examples in comparison with other notions of dissipative time crystals to begin classifying the different phenomena which can occur in open systems.}

\section*{Acknowledgements}

We thank M. Medenjak for useful discussions and collaboration on related work. We acknowledge funding from EPSRC programme grant EP/P009565/1, EPSRC National Quantum Technology Hub in Networked Quantum   Information Technology (EP/M013243/1), and the European Research Council under the European Union's Seventh Framework Programme (FP7/2007-2013)/ERC Grant Agreement no. 319286, Q-MAC.

\newpage
\section*{References}
\bibliographystyle{vancouver}
\bibliography{main}

\begin{thebibliography}{100}

\bibitem{Anderson}
Anderson PW.
\newblock More Is Different.
\newblock Science. 1972;177(4047):393--396.
\newblock Available from:
  \url{https://science.sciencemag.org/content/177/4047/393}.

\bibitem{ETHReview}
D'Alessio L, Kafri Y, Polkovnikov A, Rigol M.
\newblock From quantum chaos and eigenstate thermalization to statistical
  mechanics and thermodynamics.
\newblock Advances in Physics. 2016;65(3):239--362.
\newblock Available from: \url{https://doi.org/10.1080/00018732.2016.1198134}.

\bibitem{ETHde}
Deutsch JM.
\newblock Quantum statistical mechanics in a closed system.
\newblock Phys Rev A. 1991 Feb;43:2046--2049.
\newblock Available from:
  \url{https://link.aps.org/doi/10.1103/PhysRevA.43.2046}.

\bibitem{ETHalt}
Rigol M, Dunjko V, Olshanii M.
\newblock Thermalization and its mechanism for generic isolated quantum
  systems.
\newblock Nature. 2008 04;452:854 EP --.
\newblock Available from: \url{https://doi.org/10.1038/nature06838}.

\bibitem{eigenstatedephasing}
Barthel T, Schollw\"ock U.
\newblock Dephasing and the Steady State in Quantum Many-Particle Systems.
\newblock Phys Rev Lett. 2008 Mar;100:100601.
\newblock Available from:
  \url{https://link.aps.org/doi/10.1103/PhysRevLett.100.100601}.

\bibitem{EsslerGGE}
Essler FHL, Mussardo G, Panfil M.
\newblock Generalized Gibbs ensembles for quantum field theories.
\newblock Phys Rev A. 2015 May;91:051602(R).
\newblock Available from:
  \url{https://link.aps.org/doi/10.1103/PhysRevA.91.051602}.

\bibitem{thermoreview}
Essler FHL, Fagotti M.
\newblock Quench dynamics and relaxation in isolated integrable quantum spin
  chains.
\newblock Journal of Statistical Mechanics: Theory and Experiment. 2016
  jun;2016(6):064002.
\newblock Available from:
  \url{https://doi.org/10.1088%2F1742-5468%2F2016%2F06%2F064002}.

\bibitem{soliton}
Szankowski P, Trippenbach M, Infeld E, Rowlands G.
\newblock Oscillating Solitons in a Three-Component Bose-Einstein Condensate.
\newblock Phys Rev Lett. 2010 Sep;105:125302.
\newblock Available from:
  \url{https://link.aps.org/doi/10.1103/PhysRevLett.105.125302}.

\bibitem{Gabor}
Hódsági K, Kormos M, Takács G.
\newblock Perturbative post-quench overlaps in quantum field theory.
\newblock Journal of High Energy Physics. 2019 Aug;2019(8).
\newblock Available from: \url{http://dx.doi.org/10.1007/JHEP08(2019)047}.

\bibitem{QPWT}
Lin CJ, Motrunich OI.
\newblock Quasiparticle explanation of the weak-thermalization regime under
  quench in a nonintegrable quantum spin chain.
\newblock Phys Rev A. 2017 Feb;95:023621.
\newblock Available from:
  \url{https://link.aps.org/doi/10.1103/PhysRevA.95.023621}.

\bibitem{scars}
Turner CJ, Michailidis AA, Abanin DA, Serbyn M, Papi{\'c} Z.
\newblock Weak ergodicity breaking from quantum many-body scars.
\newblock Nature Physics. 2018;14(7):745--749.
\newblock Available from: \url{https://doi.org/10.1038/s41567-018-0137-5}.

\bibitem{scars1}
Choi S, Turner CJ, Pichler H, Ho WW, Michailidis AA,
  Papi\ifmmode~\acute{c}\else \'{c}\fi{} Z, et~al.
\newblock Emergent SU(2) Dynamics and Perfect Quantum Many-Body Scars.
\newblock Phys Rev Lett. 2019 Jun;122:220603.
\newblock Available from:
  \url{https://link.aps.org/doi/10.1103/PhysRevLett.122.220603}.

\bibitem{scars3}
Michailidis AA, Turner CJ, Papić Z, Abanin DA, Serbyn M. Stabilizing
  two-dimensional quantum scars by deformation and synchronization; 2020.

\bibitem{scarsexp}
Bernien H, Schwartz S, Keesling A, Levine H, Omran A, Pichler H, et~al.
\newblock Probing many-body dynamics on a 51-atom quantum simulator.
\newblock Nature. 2017 11;551:579 EP --.
\newblock Available from: \url{https://doi.org/10.1038/nature24622}.

\bibitem{scars4}
Lin CJ, Motrunich OI.
\newblock Exact Quantum Many-Body Scar States in the Rydberg-Blockaded Atom
  Chain.
\newblock Phys Rev Lett. 2019 Apr;122:173401.
\newblock Available from: \url{https://link.aps.org/doi/10.1103/PhysRevLet}.

\bibitem{scarsdynsim}
Bull K, Desaules JY, Papic Z. Quantum scars as embeddings of weakly "broken"
  Lie algebra representations; 2020.

\bibitem{scardynsym2}
Moudgalya S, Regnault N, Bernevig BA. Eta-Pairing in Hubbard Models: From
  Spectrum Generating Algebras to Quantum Many-Body Scars; 2020.

\bibitem{scarsdynsym3}
Mark DK, Motrunich OI. Eta-pairing states as true scars in an extended Hubbard
  Model; 2020.

\bibitem{scarsdynsym4}
Mark DK, Lin CJ, Motrunich OI. Unified structure for exact towers of scar
  states in the AKLT and other models; 2020.

\bibitem{Bojan1}
{\v{Z}}unkovi{\v{c}} B, Heyl M, Knap M, Silva A.
\newblock Dynamical quantum phase transitions in spin chains with long-range
  interactions: Merging different concepts of nonequilibrium criticality.
\newblock Physical review letters. 2018;120(13):130601.

\bibitem{Bojan2}
{\v{Z}}unkovi{\v{c}} B, Silva A, Fabrizio M.
\newblock Dynamical phase transitions and Loschmidt echo in the infinite-range
  XY model.
\newblock Philosophical Transactions of the Royal Society A: Mathematical,
  Physical and Engineering Sciences. 2016;374(2069):20150160.

\bibitem{longrange}
Zhang J, Pagano G, Hess PW, Kyprianidis A, Becker P, Kaplan H, et~al.
\newblock Observation of a many-body dynamical phase transition with a 53-qubit
  quantum simulator.
\newblock Nature. 2017;551(7682):601--604.

\bibitem{Jad1}
Zauner-Stauber V, Halimeh JC.
\newblock Probing the anomalous dynamical phase in long-range quantum spin
  chains through Fisher-zero lines.
\newblock Phys Rev E. 2017 Dec;96:062118.
\newblock Available from:
  \url{https://link.aps.org/doi/10.1103/PhysRevE.96.062118}.

\bibitem{Jad2}
Halimeh JC, Zauner-Stauber V, McCulloch IP, de~Vega I, Schollw\"ock U, Kastner
  M.
\newblock Prethermalization and persistent order in the absence of a thermal
  phase transition.
\newblock Phys Rev B. 2017 Jan;95:024302.
\newblock Available from:
  \url{https://link.aps.org/doi/10.1103/PhysRevB.95.024302}.

\bibitem{Jad3}
Halimeh JC, Zauner-Stauber V.
\newblock Dynamical phase diagram of quantum spin chains with long-range
  interactions.
\newblock Phys Rev B. 2017 Oct;96:134427.
\newblock Available from:
  \url{https://link.aps.org/doi/10.1103/PhysRevB.96.134427}.

\bibitem{Confinement5}
Kormos M, Collura M, Tak{\'a}cs G, Calabrese P.
\newblock Real-time confinement following a quantum quench to a non-integrable
  model.
\newblock Nature Physics. 2017;13(3):246--249.

\bibitem{Confinement1}
Chanda T, Zakrzewski J, Lewenstein M, Tagliacozzo L. Confinement and lack of
  thermalization after quenches in the bosonic Schwinger model; 2019.

\bibitem{Confinement2}
Lerose A, Surace FM, Mazza PP, Perfetto G, Collura M, Gambassi A.
  Quasilocalized dynamics from confinement of quantum excitations; 2019.

\bibitem{Confinement3}
Cubero AC, Robinson NJ. Lack of thermalization in (1+1)-d QCD at large $N_c$;
  2019.

\bibitem{Confinement4}
Pai S, Pretko M.
\newblock Fractons from confinement in one dimension.
\newblock Physical Review Research. 2020 Jan;2(1).
\newblock Available from:
  \url{http://dx.doi.org/10.1103/PhysRevResearch.2.013094}.

\bibitem{Confinement6}
Robinson NJ, James AJ, Konik RM.
\newblock Signatures of rare states and thermalization in a theory with
  confinement.
\newblock Physical Review B. 2019;99(19):195108.

\bibitem{breathing1}
Mittal KM, Mistakidis SI, Kevrekidis PG, Schmelcher P. Many-body effects on
  second-order phase transitions in spinor Bose-Einstein condensates and
  breathing dynamics; 2020.

\bibitem{breathing2}
Abraham JW, Balzer K, Hochstuhl D, Bonitz M.
\newblock Quantum breathing mode of interacting particles in a one-dimensional
  harmonic trap.
\newblock Phys Rev B. 2012 Sep;86:125112.
\newblock Available from:
  \url{https://link.aps.org/doi/10.1103/PhysRevB.86.125112}.

\bibitem{breathing3}
Stenger J, Inouye S, Stamper-Kurn DM, Miesner HJ, Chikkatur AP, Ketterle W.
\newblock Spin domains in ground-state Bose--Einstein condensates.
\newblock Nature. 1998;396(6709):345--348.
\newblock Available from: \url{https://doi.org/10.1038/24567}.

\bibitem{MarkoJacopo}
Medenjak M, De~Nardis J.
\newblock Domain wall melting in spin-1 XXZ chains.
\newblock Physical Review B. 2020 Feb;101(8).
\newblock Available from: \url{http://dx.doi.org/10.1103/PhysRevB.101.081411}.

\bibitem{breathing4}
Pr{\"u}fer M, Kunkel P, Strobel H, Lannig S, Linnemann D, Schmied CM, et~al.
\newblock Observation of universal dynamics in a spinor Bose gas far from
  equilibrium.
\newblock Nature. 2018;563(7730):217--220.

\bibitem{breathing5}
Adams A, Carr LD, Schaefer T, Steinberg P, Thomas JE.
\newblock Strongly correlated quantum fluids: ultracold quantum gases, quantum
  chromodynamic plasmas and holographic duality.
\newblock New Journal of Physics. 2012;14(11):115009.

\bibitem{breathing6}
Fang B, Carleo G, Johnson A, Bouchoule I.
\newblock Quench-Induced Breathing Mode of One-Dimensional Bose Gases.
\newblock Physical Review Letters. 2014 Jul;113(3).
\newblock Available from:
  \url{http://dx.doi.org/10.1103/PhysRevLett.113.035301}.

\bibitem{randommatrix1}
Denisov S, Laptyeva T, Tarnowski W, Chru\ifmmode \acute{s}\else
  \'{s}\fi{}ci\ifmmode~\acute{n}\else \'{n}\fi{}ski D, \ifmmode~\dot{Z}\else
  \.~{Z}\fi{}yczkowski K.
\newblock Universal Spectra of Random Lindblad Operators.
\newblock Phys Rev Lett. 2019 Oct;123:140403.
\newblock Available from:
  \url{https://link.aps.org/doi/10.1103/PhysRevLett.123.140403}.

\bibitem{randommatrix2}
Wang K, Piazza F, Luitz DJ.
\newblock Hierarchy of Relaxation Timescales in Local Random Liouvillians.
\newblock Phys Rev Lett. 2020 Mar;124:100604.
\newblock Available from:
  \url{https://link.aps.org/doi/10.1103/PhysRevLett.124.100604}.

\bibitem{randommatrix3}
Can T.
\newblock Random Lindblad dynamics.
\newblock Journal of Physics A: Mathematical and Theoretical. 2019
  nov;52(48):485302.
\newblock Available from: \url{https://doi.org/10.1088%2F1751-8121%2Fab4d26}.

\bibitem{randommatrix4}
Sá L, Ribeiro P, Prosen T. Spectral and Steady-State Properties of Random
  Liouvillians; 2019.

\bibitem{randommatrix5}
Can T, Oganesyan V, Orgad D, Gopalakrishnan S.
\newblock Spectral Gaps and Midgap States in Random Quantum Master Equations.
\newblock Phys Rev Lett. 2019 Dec;123:234103.
\newblock Available from:
  \url{https://link.aps.org/doi/10.1103/PhysRevLett.123.234103}.

\bibitem{BaumNarn}
Baumgartner B, Narnhofer H.
\newblock Analysis of quantum semigroups with GKS--Lindblad generators: II.
  General.
\newblock Journal of Physics A: Mathematical and Theoretical.
  2008;41(39):395303.

\bibitem{AlbertJiang}
Albert VV, Jiang L.
\newblock Symmetries and conserved quantities in Lindblad master equations.
\newblock Physical Review A. 2014;89(2):022118.

\bibitem{ChoiNoiselesssubsystems}
Choi MD, Kribs DW.
\newblock Method to Find Quantum Noiseless Subsystems.
\newblock Phys Rev Lett. 2006 Feb;96:050501.
\newblock Available from:
  \url{https://link.aps.org/doi/10.1103/PhysRevLett.96.050501}.

\bibitem{buca_non-stationary_2019}
Buča B, Tindall J, Jaksch D.
\newblock Non-stationary coherent quantum many-body dynamics through
  dissipation.
\newblock Nature Communications. 2019 Dec;10(1):1730.

\bibitem{Yi_2012}
Yi W, Diehl S, Daley AJ, Zoller P.
\newblock Driven-dissipative many-body pairing states for cold fermionic atoms
  in an optical lattice.
\newblock New Journal of Physics. 2012 may;14(5):055002.
\newblock Available from:
  \url{https://doi.org/10.1088%2F1367-2630%2F14%2F5%2F055002}.

\bibitem{Wilczek}
Wilczek F.
\newblock Quantum Time Crystals.
\newblock Phys Rev Lett. 2012 Oct;109:160401.
\newblock Available from:
  \url{https://link.aps.org/doi/10.1103/PhysRevLett.109.160401}.

\bibitem{bruno1}
Bruno P.
\newblock Comment on ``Quantum Time Crystals''.
\newblock Phys Rev Lett. 2013 Mar;110:118901.
\newblock Available from:
  \url{https://link.aps.org/doi/10.1103/PhysRevLett.110.118901}.

\bibitem{bruno2}
Bruno P.
\newblock Comment on ``Space-Time Crystals of Trapped Ions''.
\newblock Phys Rev Lett. 2013 Jul;111:029301.
\newblock Available from:
  \url{https://link.aps.org/doi/10.1103/PhysRevLett.111.029301}.

\bibitem{watanabe}
Watanabe H, Oshikawa M.
\newblock Absence of Quantum Time Crystals.
\newblock Phys Rev Lett. 2015 Jun;114:251603.
\newblock Available from:
  \url{https://link.aps.org/doi/10.1103/PhysRevLett.114.251603}.

\bibitem{Marko1}
Medenjak M, Buca B, Jaksch D. The isolated Heisenberg magnet as a quantum time
  crystal; 2019.

\bibitem{Cardybook}
Cardy J.
\newblock Scaling and renormalization in statistical physics. vol.~5.
\newblock Cambridge university press; 1996.

\bibitem{Marko2}
Medenjak M, Prosen T, Zadnik L. Rigorous bounds on dynamical response functions
  and time-translation symmetry breaking; 2020.

\bibitem{chinzei2020time}
Chinzei K, Ikeda TN. Time Crystals Protected by Floquet Dynamical Symmetry in
  Hubbard Models; 2020.

\bibitem{VidmarRigol}
Vidmar L, Rigol M.
\newblock Generalized Gibbs ensemble in integrable lattice models.
\newblock Journal of Statistical Mechanics: Theory and Experiment. 2016
  jun;2016(6):064007.
\newblock Available from:
  \url{https://doi.org/10.1088%2F1742-5468%2F2016%2F06%2F064007}.

\bibitem{VedikaReview}
{Khemani} V, {Moessner} R, {Sondhi} SL.
\newblock {A Brief History of Time Crystals}.
\newblock arXiv e-prints. 2019 Oct;p. arXiv:1910.10745.

\bibitem{QuantumSynch}
Tindall J, Mu{\~{n}}oz CS, Bu{\v{c}}a B, Jaksch D.
\newblock Quantum synchronisation enabled by dynamical symmetries and
  dissipation.
\newblock New Journal of Physics. 2020 jan;22(1):013026.
\newblock Available from: \url{https://doi.org/10.1088%2F1367-2630%2Fab60f5}.

\bibitem{QuantSynch1}
Eneriz H, Rossatto DZ, C{\'a}rdenas-L{\'o}pez FA, Solano E, Sanz M.
\newblock Degree of Quantumness in Quantum Synchronization.
\newblock Scientific Reports. 2019;9(1):19933.
\newblock Available from: \url{https://doi.org/10.1038/s41598-019-56468-x}.

\bibitem{QuantSynch2}
Cabot A, Giorgi GL, Zambrini R. Synchronization and coalescence in a
  dissipative two-qubit system; 2019.

\bibitem{VedikaLazarides}
Khemani V, Lazarides A, Moessner R, Sondhi SL.
\newblock Phase Structure of Driven Quantum Systems.
\newblock Phys Rev Lett. 2016 Jun;116:250401.
\newblock Available from:
  \url{https://link.aps.org/doi/10.1103/PhysRevLett.116.250401}.

\bibitem{timecrystal1}
Else DV, Bauer B, Nayak C.
\newblock Floquet Time Crystals.
\newblock Phys Rev Lett. 2016 Aug;117:090402.
\newblock Available from:
  \url{https://link.aps.org/doi/10.1103/PhysRevLett.117.090402}.

\bibitem{timecrystal2}
Lazarides A, Das A, Moessner R.
\newblock Periodic Thermodynamics of Isolated Quantum Systems.
\newblock Phys Rev Lett. 2014 Apr;112:150401.
\newblock Available from:
  \url{https://link.aps.org/doi/10.1103/PhysRevLett.112.150401}.

\bibitem{timecrystal4}
Sacha K, Zakrzewski J.
\newblock Time crystals: a review.
\newblock Reports on Progress in Physics. 2017 nov;81(1):016401.
\newblock Available from: \url{https://doi.org/10.1088%2F1361-6633%2Faa8b38}.

\bibitem{timecrystal5}
Yu WC, Tangpanitanon J, Glaetzle AW, Jaksch D, Angelakis DG.
\newblock Discrete time crystal in globally driven interacting quantum systems
  without disorder.
\newblock Phys Rev A. 2019 Mar;99:033618.
\newblock Available from:
  \url{https://link.aps.org/doi/10.1103/PhysRevA.99.033618}.

\bibitem{timecrystal7}
{Gambetta} FM, {Carollo} F, {Lazarides} A, {Lesanovsky} I, {Garrahan} JP.
\newblock {Classical Stochastic Discrete Time Crystals}.
\newblock arXiv e-prints. 2019 May;p. arXiv:1905.08826.

\bibitem{timecrystal8}
Giergiel K, Dauphin A, Lewenstein M, Zakrzewski J, Sacha K.
\newblock Topological time crystals.
\newblock New Journal of Physics. 2019 may;21(5):052003.
\newblock Available from: \url{https://doi.org/10.1088%2F1367-2630%2Fab1e5f}.

\bibitem{timecrystal10}
Sacha K, Zakrzewski J.
\newblock Time crystals: a review.
\newblock Reports on Progress in Physics. 2017;81(1):016401.

\bibitem{timecrystal11}
Ivanov DA, Ivanova TY, Caballero-Benitez SF, Mekhov IB.
\newblock Feedback-Induced Quantum Phase Transitions Using Weak Measurements.
\newblock Phys Rev Lett. 2020 Jan;124:010603.
\newblock Available from:
  \url{https://link.aps.org/doi/10.1103/PhysRevLett.124.010603}.

\bibitem{timecrystal12}
Homann G, Cosme JG, Mathey L. Higgs Time Crystal in a High-$T_{\mathrm{c}}$
  Superconductor; 2020.

\bibitem{timecrystal6}
Zhu B, Marino J, Yao NY, Lukin MD, Demler EA.
\newblock Dicke time crystals in driven-dissipative quantum many-body systems.
\newblock arXiv preprint arXiv:190401026. 2019;.

\bibitem{timecrystal9}
Gambetta FM, Carollo F, Marcuzzi M, Garrahan JP, Lesanovsky I.
\newblock Discrete Time Crystals in the Absence of Manifest Symmetries or
  Disorder in Open Quantum Systems.
\newblock Phys Rev Lett. 2019 Jan;122:015701.
\newblock Available from:
  \url{https://link.aps.org/doi/10.1103/PhysRevLett.122.015701}.

\bibitem{LazaridesRoy}
Lazarides A, Roy S, Piazza F, Moessner R.
\newblock Time crystallinity in dissipative Floquet systems.
\newblock Phys Rev Research. 2020 Apr;2:022002.
\newblock Available from:
  \url{https://link.aps.org/doi/10.1103/PhysRevResearch.2.022002}.

\bibitem{timecrystal13}
Wang RRW, Xing B, Carlo GG, Poletti D.
\newblock Period doubling in period-one steady states.
\newblock Phys Rev E. 2018 Feb;97:020202.
\newblock Available from:
  \url{https://link.aps.org/doi/10.1103/PhysRevE.97.020202}.

\bibitem{classical1}
Gambetta FM, Carollo F, Lazarides A, Lesanovsky I, Garrahan JP.
\newblock Classical stochastic discrete time crystals.
\newblock Physical Review E. 2019 Dec;100(6).
\newblock Available from: \url{http://dx.doi.org/10.1103/PhysRevE.100.060105}.

\bibitem{classical3}
Heugel TL, Oscity M, Eichler A, Zilberberg O, Chitra R.
\newblock Classical many-body time crystals.
\newblock Physical review letters. 2019;123(12):124301.

\bibitem{classical4}
Fruchart M, Hanai R, Littlewood PB, Vitelli V.
\newblock Phase transitions in non-reciprocal active systems.
\newblock arXiv preprint arXiv:200313176. 2020;.

\bibitem{Pablo}
Hurtado-Gutiérrez R, Carollo F, Pérez-Espigares C, Hurtado PI. Building
  continuous time crystals from rare events; 2019.

\bibitem{kozin2019}
Kozin VK, Kyriienko O.
\newblock Quantum Time Crystals from Hamiltonians with Long-Range Interactions.
\newblock Phys Rev Lett. 2019 Nov;123:210602.
\newblock Available from:
  \url{https://link.aps.org/doi/10.1103/PhysRevLett.123.210602}.

\bibitem{quasicrystal1}
Pizzi A, Knolle J, Nunnenkamp A.
\newblock Period-$n$ Discrete Time Crystals and Quasicrystals with Ultracold
  Bosons.
\newblock Phys Rev Lett. 2019 Oct;123:150601.
\newblock Available from:
  \url{https://link.aps.org/doi/10.1103/PhysRevLett.123.150601}.

\bibitem{Sarang}
Chan CK, Lee TE, Gopalakrishnan S.
\newblock Limit-cycle phase in driven-dissipative spin systems.
\newblock Phys Rev A. 2015 May;91:051601.
\newblock Available from:
  \url{https://link.aps.org/doi/10.1103/PhysRevA.91.051601}.

\bibitem{disTC}
Chiacchio EIR, Nunnenkamp A.
\newblock Dissipation-Induced Instabilities of a Spinor Bose-Einstein
  Condensate Inside an Optical Cavity.
\newblock Phys Rev Lett. 2019 May;122:193605.
\newblock Available from:
  \url{https://link.aps.org/doi/10.1103/PhysRevLett.122.193605}.

\bibitem{disTC2}
Barberena D, Lewis-Swan RJ, Thompson JK, Rey AM.
\newblock Driven-dissipative quantum dynamics in ultra-long-lived dipoles in an
  optical cavity.
\newblock Phys Rev A. 2019 May;99:053411.
\newblock Available from:
  \url{https://link.aps.org/doi/10.1103/PhysRevA.99.053411}.

\bibitem{disTC3}
Lled\'o C, Mavrogordatos TK, Szyma\ifmmode~\acute{n}\else \'{n}\fi{}ska MH.
\newblock Driven Bose-Hubbard dimer under nonlocal dissipation: A bistable time
  crystal.
\newblock Phys Rev B. 2019 Aug;100:054303.
\newblock Available from:
  \url{https://link.aps.org/doi/10.1103/PhysRevB.100.054303}.

\bibitem{disTC4}
Cosme JG, Skulte J, Mathey L.
\newblock Time crystals in a shaken atom-cavity system.
\newblock arXiv preprint arXiv:190900266. 2019;.

\bibitem{Esslinger}
Dogra N, Landini M, Kroeger K, Hruby L, Donner T, Esslinger T.
\newblock Dissipation-induced structural instability and chiral dynamics in a
  quantum gas.
\newblock Science. 2019;366(6472):1496--1499.
\newblock Available from:
  \url{https://science.sciencemag.org/content/366/6472/1496}.

\bibitem{Esslinger2}
Zupancic P, Dreon D, Li X, Baumg\"artner A, Morales A, Zheng W, et~al.
\newblock $P$-Band Induced Self-Organization and Dynamics with Repulsively
  Driven Ultracold Atoms in an Optical Cavity.
\newblock Phys Rev Lett. 2019 Dec;123:233601.
\newblock Available from:
  \url{https://link.aps.org/doi/10.1103/PhysRevLett.123.233601}.

\bibitem{BucaJaksch2019}
Bu\ifmmode~\check{c}\else \v{c}\fi{}a B, Jaksch D.
\newblock Dissipation Induced Nonstationarity in a Quantum Gas.
\newblock Phys Rev Lett. 2019 Dec;123:260401.
\newblock Available from:
  \url{https://link.aps.org/doi/10.1103/PhysRevLett.123.260401}.

\bibitem{disTC5}
Seibold K, Rota R, Savona V.
\newblock Dissipative time crystal in an asymmetric nonlinear photonic dimer.
\newblock Phys Rev A. 2020 Mar;101:033839.
\newblock Available from:
  \url{https://link.aps.org/doi/10.1103/PhysRevA.101.033839}.

\bibitem{disTC6}
Riera-Campeny A, Moreno-Cardoner M, Sanpera A. Time crystallinity in open
  quantum systems; 2019.

\bibitem{disTC7}
Tucker K, Zhu B, Lewis-Swan RJ, Marino J, Jimenez F, Restrepo JG, et~al.
\newblock Shattered time: can a dissipative time crystal survive many-body
  correlations?
\newblock New Journal of Physics. 2018;20(12):123003.

\bibitem{disTC8}
Scarlatella O, Fazio R, Schir\'o M.
\newblock Emergent finite frequency criticality of driven-dissipative
  correlated lattice bosons.
\newblock Phys Rev B. 2019 Feb;99:064511.
\newblock Available from:
  \url{https://link.aps.org/doi/10.1103/PhysRevB.99.064511}.

\bibitem{disTC9}
Ke\ss{}ler H, Cosme JG, Hemmerling M, Mathey L, Hemmerich A.
\newblock Emergent limit cycles and time crystal dynamics in an atom-cavity
  system.
\newblock Phys Rev A. 2019 May;99:053605.
\newblock Available from:
  \url{https://link.aps.org/doi/10.1103/PhysRevA.99.053605}.

\bibitem{disTC10}
Marconi M, Raineri F, Levenson A, Yacomotti AM, Javaloyes J, Pan SH, et~al.
\newblock Mesoscopic limit cycles in coupled nanolasers.
\newblock arXiv preprint arXiv:191110830. 2019;.

\bibitem{disTC11}
Lled{\'o} C, Szyma{\'n}ska MH.
\newblock A dissipative time crystal with or without $\cal{Z}_2$ symmetry
  breaking.
\newblock arXiv preprint arXiv:200402855. 2020;.

\bibitem{disTC12}
K\'onya G, Nagy D, Szirmai G, Domokos P.
\newblock Nonequilibrium polariton dynamics in a Bose-Einstein condensate
  coupled to an optical cavity.
\newblock Phys Rev A. 2018 Dec;98:063608.
\newblock Available from:
  \url{https://link.aps.org/doi/10.1103/PhysRevA.98.063608}.

\bibitem{disTC13}
Soriente M, Donner T, Chitra R, Zilberberg O.
\newblock Dissipation-Induced Anomalous Multicritical Phenomena.
\newblock Phys Rev Lett. 2018 May;120:183603.
\newblock Available from:
  \url{https://link.aps.org/doi/10.1103/PhysRevLett.120.183603}.

\bibitem{disTC14}
Hongo M, Kim S, Noumi T, Ota A.
\newblock Effective field theory of time-translational symmetry breaking in
  nonequilibrium open system.
\newblock Journal of High Energy Physics. 2019;2019(2):131.

\bibitem{disTC15}
Fan J, Chen G, Jia S. Atomic self-organization emerging from a tunable
  quadratures coupling; 2020.

\bibitem{disTC16}
Keßler H, Cosme JG, Georges C, Mathey L, Hemmerich A. From a continuous to a
  discrete time crystal; 2020.

\bibitem{Fazio}
Iemini F, Russomanno A, Keeling J, Schir\`o M, Dalmonte M, Fazio R.
\newblock Boundary Time Crystals.
\newblock Phys Rev Lett. 2018 Jul;121:035301.
\newblock Available from:
  \url{https://link.aps.org/doi/10.1103/PhysRevLett.121.035301}.

\bibitem{boundary2}
Muñoz CS, Buča B, Tindall J, González-Tudela A, Jaksch D, Porras D.
  Non-stationary dynamics and dissipative freezing in squeezed superradiance;
  2019.

\bibitem{speckle}
Schreiber M, Hodgman SS, Bordia P, L{\"u}schen HP, Fischer MH, Vosk R, et~al.
\newblock Observation of many-body localization of interacting fermions in a
  quasirandom optical lattice.
\newblock Science. 2015;349(6250):842--845.

\bibitem{Gorin}
Gorin T, Prosen T, Seligman TH, {\v{Z}}nidari{\v{c}} M.
\newblock Dynamics of Loschmidt echoes and fidelity decay.
\newblock Physics Reports. 2006;435(2-5):33--156.

\bibitem{DPTreview}
Heyl M.
\newblock Dynamical quantum phase transitions: a review.
\newblock Reports on Progress in Physics. 2018;81(5):054001.

\bibitem{Zanardi}
Campos~Venuti L, Zanardi P.
\newblock Unitary equilibrations: Probability distribution of the Loschmidt
  echo.
\newblock Phys Rev A. 2010 Feb;81:022113.
\newblock Available from:
  \url{https://link.aps.org/doi/10.1103/PhysRevA.81.022113}.

\bibitem{Zanardi2}
Campos~Venuti L, Jacobson NT, Santra S, Zanardi P.
\newblock Exact Infinite-Time Statistics of the Loschmidt Echo for a Quantum
  Quench.
\newblock Phys Rev Lett. 2011 Jul;107:010403.
\newblock Available from:
  \url{https://link.aps.org/doi/10.1103/PhysRevLett.107.010403}.

\bibitem{Echo1}
Häppölä J, Halász GB, Hamma A.
\newblock Universality and robustness of revivals in the transverse
  fieldXYmodel.
\newblock Physical Review A. 2012 Mar;85(3).
\newblock Available from: \url{http://dx.doi.org/10.1103/PhysRevA.85.032114}.

\bibitem{lindblad1976}
Lindblad G.
\newblock On the generators of quantum dynamical semigroups.
\newblock Comm Math Phys. 1976;(2):119--130.

\bibitem{BPTextbook}
Breuer HP, Petruccione F, et~al.
\newblock The theory of open quantum systems.
\newblock Oxford University Press on Demand; 2002.

\bibitem{GardinerTextbook}
Gardiner C, Zoller P, Zoller P.
\newblock Quantum noise: a handbook of Markovian and non-Markovian quantum
  stochastic methods with applications to quantum optics.
\newblock Springer Science \& Business Media; 2004.

\bibitem{limitcycle1}
Hush MR, Li W, Genway S, Lesanovsky I, Armour AD.
\newblock Spin correlations as a probe of quantum synchronization in
  trapped-ion phonon lasers.
\newblock Physical Review A. 2015;91(6):061401.

\bibitem{limitcycle2}
Rotondo P, Marcuzzi M, Garrahan JP, Lesanovsky I, M{\"u}ller M.
\newblock Open quantum generalisation of Hopfield neural networks.
\newblock Journal of Physics A: Mathematical and Theoretical.
  2018;51(11):115301.

\bibitem{Lidar}
Lidar DA, Chuang IL, Whaley KB.
\newblock Decoherence-free subspaces for quantum computation.
\newblock Physical Review Letters. 1998;81(12):2594.

\bibitem{Albert}
Albert VV, Bradlyn B, Fraas M, Jiang L.
\newblock Geometry and response of Lindbladians.
\newblock Physical Review X. 2016;6(4):041031.

\bibitem{BucaProsen}
Bu{\v{c}}a B, Prosen T.
\newblock A note on symmetry reductions of the Lindblad equation: transport in
  constrained open spin chains.
\newblock New Journal of Physics. 2012 jul;14(7):073007.
\newblock Available from:
  \url{https://doi.org/10.1088%2F1367-2630%2F14%2F7%2F073007}.

\bibitem{Zhao}
Zhang Z, Tindall J, Mur-Petit J, Jaksch D, Buca B.
\newblock Stationary state degeneracy of open quantum systems with non-Abelian
  symmetries.
\newblock Journal of Physics A: Mathematical and Theoretical. 2020;.

\bibitem{coldexp}
Gross C, Bloch I.
\newblock Quantum simulations with ultracold atoms in optical lattices.
\newblock Science. 2017;357(6355):995--1001.
\newblock Available from:
  \url{https://science.sciencemag.org/content/357/6355/995}.

\bibitem{fftwindowing}
{Harris} FJ.
\newblock On the use of windows for harmonic analysis with the discrete Fourier
  transform.
\newblock Proceedings of the IEEE. 1978;66(1):51--83.

\bibitem{essler_frahm_gohmann_klumper_korepin_2005}
Essler FHL, Frahm H, Göhmann F, Klümper A, Korepin VE.
\newblock The One-Dimensional Hubbard Model.
\newblock Cambridge University Press; 2005.

\bibitem{howard_signal_2015}
Howard RM.
\newblock A {Signal} {Theoretic} {Introduction} to {Random} {Processes}.
\newblock Hoboken, UNITED STATES: John Wiley \& Sons, Incorporated; 2015.
\newblock Available from:
  \url{http://ebookcentral.proquest.com/lib/oxford/detail.action?docID=1895995}.

\bibitem{buca_non-stationary_2019_sup}
Buča B, Tindall J, Jaksch D.
\newblock Non-stationary coherent quantum many-body dynamics through
  dissipation - Supplimentary material.
\newblock Nature Communications. 2019 Dec;10(1):1730.

\bibitem{quasicrystal2}
Giergiel K, Kuro\ifmmode~\acute{s}\else \'{s}\fi{} A, Sacha K.
\newblock Discrete time quasicrystals.
\newblock Phys Rev B. 2019 Jun;99:220303.
\newblock Available from:
  \url{https://link.aps.org/doi/10.1103/PhysRevB.99.220303}.

\bibitem{DaleyQuantumTrajectories}
Daley AJ.
\newblock Quantum trajectories and open many-body quantum systems.
\newblock Advances in Physics. 2014;63(2):77--149.
\newblock Available from: \url{https://doi.org/10.1080/00018732.2014.933502}.

\bibitem{Nurdin_2020}
Nurdin HI.
\newblock Quantum Stochastic Processes and the Modelling of Quantum Noise.
\newblock Encyclopedia of Systems and Control. 2020;p. 1–8.
\newblock Available from:
  \url{http://dx.doi.org/10.1007/978-1-4471-5102-9_100160-1}.

\bibitem{Vedikacomment}
Khemani V, Moessner R, Sondhi SL. Comment on "Quantum Time Crystals from
  Hamiltonians with Long-Range Interactions"; 2020.

\bibitem{Eisert}
Kshetrimayum A, Goihl M, Kennes DM, Eisert J. Quantum time crystals with
  programmable disorder in higher dimensions; 2020.

\bibitem{Lieb_Robinson_1972}
Lieb EH, Robinson DW.
\newblock The finite group velocity of quantum spin systems.
\newblock Communications in Mathematical Physics. 1972 Sep;28(3):251–257.

\bibitem{Znidaric_2015}
Žnidarič M.
\newblock Relaxation times of dissipative many-body quantum systems.
\newblock Physical Review E. 2015 Oct;92(4):042143.

\bibitem{Andreas}
Garcia-Mata I, Carvalho AR, Mintert F, Buchleitner A.
\newblock Entanglement screening by nonlinear resonances.
\newblock Physical review letters. 2007;98(12):120504.

\bibitem{Castro}
Castro-Alvaredo OA, Lencsés M, Szécsényi IM, Viti J. Entanglement
  Oscillations near a Quantum Critical Point; 2020.

\bibitem{Delfino}
Delfino G.
\newblock Quantum quenches with integrable pre-quench dynamics.
\newblock Journal of Physics A: Mathematical and Theoretical.
  2014;47(40):402001.

\bibitem{Luttinger1}
B{\'a}csi {\'A}, Moca CP, D{\'o}ra B.
\newblock Dissipation-Induced Luttinger Liquid Correlations in a
  One-Dimensional Fermi Gas.
\newblock Physical Review Letters. 2020;124(13):136401.

\bibitem{Luttinger2}
Langmann E, Lebowitz JL, Mastropietro V, Moosavi P.
\newblock Steady States and Universal Conductance in a Quenched Luttinger
  Model.
\newblock Communications in Mathematical Physics. 2017;349(2):551--582.
\newblock Available from: \url{https://doi.org/10.1007/s00220-016-2631-x}.

\end{thebibliography}

\end{document}